\documentclass[prb,twocolumn,floats,10pt,aps,superscriptaddress]{revtex4-2}
\usepackage{dcolumn}
\usepackage{amsmath}
\usepackage{amsfonts}
\usepackage{braket}
\usepackage{xcolor}
\usepackage{appendix}
\usepackage{enumitem}
\usepackage{graphicx}
\usepackage{float}
\usepackage{tikz}
\usepackage{algorithm}
\usepackage{algpseudocode}
\usepackage{bbm}

\RequirePackage[hyperindex,colorlinks,bookmarksnumbered, plainpages=true,pdfstartview=FitH]{hyperref}
\hypersetup{linkcolor=blue,urlcolor=blue,citecolor=blue}
\usepackage{hyperref}

\begin{document}
\title{U(1)-symmetric Gaussian fermionic projected entangled paired states and their Gutzwiller projection}

\author{Jheng-Wei Li}
\affiliation{Arnold Sommerfeld Center for Theoretical Physics, Center for NanoScience,\looseness=-1\,  and Munich
Center for \\ Quantum Science and Technology,\looseness=-2\, Ludwig-Maximilians-Universität München, 80333 Munich, Germany}

\author{Jan von Delft}
\affiliation{Arnold Sommerfeld Center for Theoretical Physics, Center for NanoScience,\looseness=-1\,  and Munich
Center for \\ Quantum Science and Technology,\looseness=-2\, Ludwig-Maximilians-Universität München, 80333 Munich, Germany}

\author{Hong-Hao Tu}
\affiliation{Institut f\"ur Theoretische Physik, Technische Universit\"at Dresden, 01062 Dresden, Germany}

\begin{abstract}

We develop a formalism for constructing particle-number-conserving Gaussian fermionic projected entangled pair states [U(1)-GfPEPS] and show that these states can describe ground states of band insulators and gapless fermions with band touching points.
When using them as variational Ans\"{a}tze for two Dirac fermion systems ($\pi$-flux model on the square lattice and $[0,\pi]$-flux model on the kagome lattice), we find that the U(1)-GfPEPS, even with a relatively small bond dimension, can accurately approximate the Dirac Fermi sea ground states. 
By applying Gutzwiller projectors on top of these U(1)-GfPEPS, we obtain PEPS representation of U(1)-Dirac spin liquid states for spin-1/2 systems.
With state-of-the-art tensor network numerics, the critical exponent in the spin-spin correlation function of the Gutzwiller-projected $\pi$-flux state is estimated to be $\eta \approx 1.7$.

\end{abstract}
\date{\today}
\maketitle

\section{Introduction}
The idea of the Gutzwiller wave function plays a crucial role in the study of strongly correlated systems.
Its original formulation considers a Slater determinant wave function for electrons and supplements that with a Gutzwiller operator accounting for electron correlations~\cite{Gutzwiller1963,Gutzwiller1965}.
Since its invention, the scope of the Gutzwiller wave function has been considerably broadened. 
For instance, Anderson has proposed a Gutzwiller projected BCS state for high-$T_c$ cuprates~\cite{Anderson1987}.
In the modern context, the Gutzwiller wave function evolves into the framework of a systematic approach called ``parton construction'', which includes three main steps: 
(i) the constituent particles (fermions, bosons, or spins) of an interacting system are split into fermionic or bosonic ``partons'' with enlarged Hilbert spaces; 
(ii) the fermionic or bosonic partons are placed into certain non-interacting (quadratic) mean-field Hamiltonians with fermionic or bosonic Gaussian ground states; 
(iii) the Gutzwiller projection, taking the form of a local projector, is applied to Gaussian ground states of partons to remove unphysical states introduced by the parton construction.
For paradigmatic examples like the Haldane-Shastry model~\cite{Haldane1988,Shastry1988} and the Kitaev's honeycomb model~\cite{Kitaev2006}, Gutzwiller wave functions are exact ground states and provide invaluable insight into exotic states emerging from strong correlations.

From a numerical perspective, the variational Monte Carlo method using Gutzwiller projected fermionic wave functions has been one of the key methods for strongly correlated systems~\cite{Gros1987,Yokoyama1987,Gros1989}. 
Recently, several methods have been developed for converting fermionic Gaussian states into matrix product states (MPSs)~\cite{Fishman2015,Wu2020,Jin2020,Aghaei2020,Petrica2021,Jones2021a,Jin2022a}.
In the MPS representation, the Gutzwiller projection can be implemented easily.
This provides not just a new approach for evaluating physical quantities in Gutzwiller wave functions, but also physically motivated MPSs for initializing density matrix renormalization group (DMRG) calculations~\cite{White1992,Ostlund1995,Verstraete2008,Schollwoeck2011}.
Such a strategy has already seen success in accelerating DMRG calculations and, for topologically ordered phases, targeting degenerate ground states in different topological sectors~\cite{Jin2021,Chen2021,Jin2022b,Sun2022}.

For two-dimensional (2D) systems, too, it is highly desirable to develop a method converting Gutzwiller projected wave functions into projected entangled pair states (PEPSs)~\cite{Verstraete2004}. 
Similar to the benefits for DMRG, Gutzwiller wave functions can serve as good initial inputs in PEPS-based variational methods~\cite{Jiang2008,Jordan2008,Corboz2016,Vanderstraeten2016,Liao2019}.
For concrete Hamiltonians, the comparison of Gutzwiller wave functions with brute-force PEPS numerical results would also become possible.
Furthermore, for 2D systems, the PEPS representation of Gutzwiller wave functions has two advantages over its MPS counterpart: 
(i) infinite-size PEPS algorithms~\cite{Nishino1996,Levin2007,Orus2009,Xie2012,Fishman2018} work directly in the thermodynamic limit, whereas the MPS approach using a cylindrical boundary condition suffers from finite-size effects; 
(ii) for topological systems, the local tensor of PEPS usually exhibits a symmetry~\cite{Schuch2010,Buerschaper2014,Williamson2016,Sahinouglu2021}, which can be used to characterize topological properties.

In this work, we develop a systematic approach to convert Gutzwiller projected Fermi sea states into PEPSs.
This is based on a specification of the Gaussian fermionic PEPS (GfPEPS) formalism~\cite{Kraus2010} to a particle-number-conserving setting (referred to as U(1)-GfPEPS hereafter).
We show that the U(1)-GfPEPS can describe band insulators whose filled valence bands and empty conduction bands are separated by a gap, as well as semimetals with band-touching points (e.g., Dirac points) between valence and conduction bands.
The case of an open Fermi surface is beyond the scope of U(1)-GfPEPS.
Furthermore, we develop a variational algorithm that starts with a particle-number-conserving free fermionic Hamiltonian and approximates its ground state with U(1)-GfPEPS.
This complements previous works focusing on analytical constructions~\cite{Wahl2013,Dubail2015,Poilblanc2014,Wahl2014,Yang2015,Hackenbroich2020} and a related numerical work which does not impose particle-number conservation~\cite{Mortier2020}.
For two Dirac fermion systems ($\pi$-flux model on the square lattice and $[0,\pi]$-flux model on the kagome lattice), the benchmark calculations with U(1)-GfPEPS accurately reproduce the filled band dispersions with a relatively small bond dimension.
The application of additional Gutzwiller projectors to these U(1)-GfPEPS provides PEPS Ans\"atze for U(1)-Dirac spin liquids.
From these we calculate their spin-spin correlation functions with state-of-the-art tensor network algorithms and obtain a critical exponent $\eta \approx 1.7$ for the Gutzwiller-projected $\pi$-flux state.

The rest of this paper is organized as follows.
In Sec.~\ref{sec:methods} we describe our methods, including the construction of U(1)-GfPEPS and its correlation matrix formalism, the variational optimization algorithm for U(1)-GfPEPS, the implementation of Gutzwiller projection, and the contraction method for computing physical quantities.
In Sec.~\ref{sec:results}, we apply these methods to two benchmark examples, i.e., the $\pi$-flux model on the square lattice and the $[0,\pi]$-flux model on the kagome lattice.
The U(1)-Dirac spin liquid states obtained after Gutzwiller projection are also studied.
Sec.~\ref{sec:summary} provides a summary and gives some outlook.
Appendix~\ref{app:U1-Gaussian-formalism} includes technical details on particle-number-conserving fermionic Gaussian states.

\section{Methods}
\label{sec:methods}

\subsection{U(1)-symmetric Gaussian fermionic projected entangled-paired state}

We use the square lattice to illustrate the construction of U(1)-GfPEPS; the extension to other lattices is straightforward. 
Each site of the lattice hosts $P$ physical fermionic modes, with creation operators $c^{\dag}_{\mathbf{r},\mu}$ ($\mu=1,\ldots,P$), as well as $4M$ virtual fermionic modes, with creation operators $c^{\dag}_{\mathbf{r},\nu,\alpha}$ ($\nu=l,r,d,u$ and $\alpha=1,\ldots,M$), where $l,r,d,u$ denote left, right, down, and up, respectively.

\begin{figure}[!htb]
  \includegraphics[width=1.0\linewidth,trim = 0.0in 0.0in 0.0in 0.0in,clip=true]{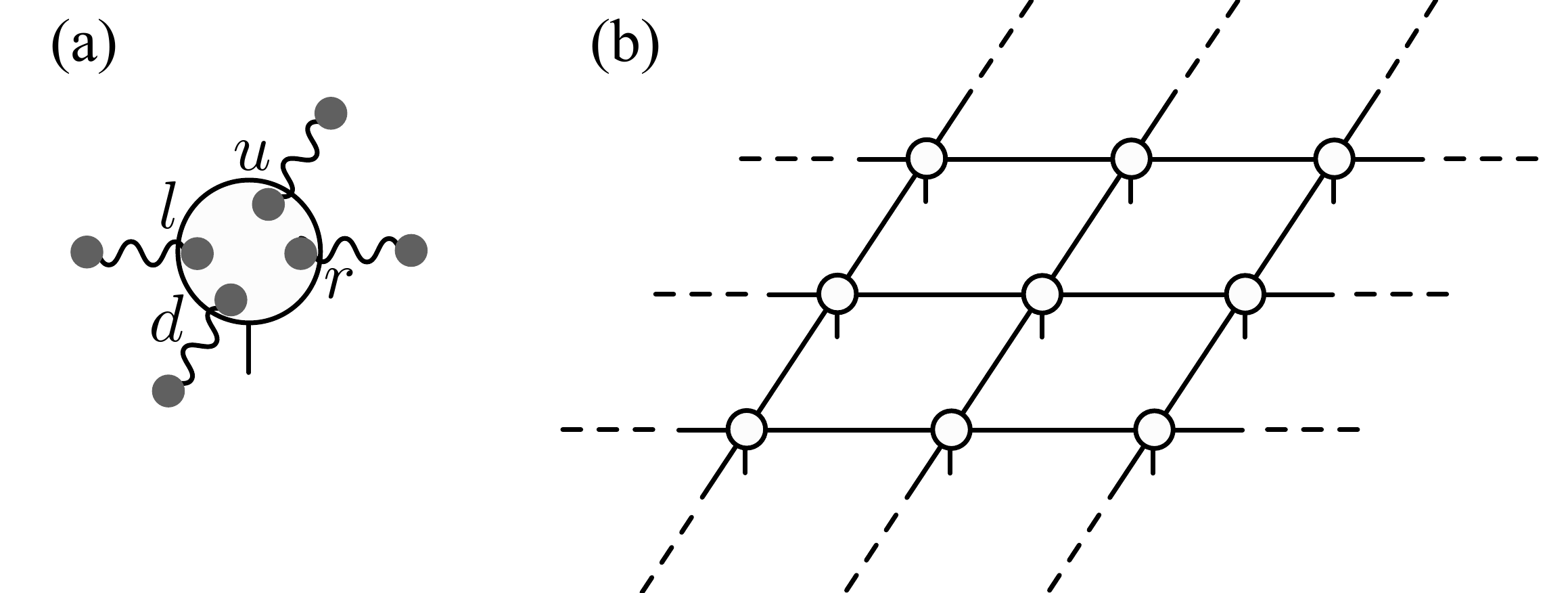}
  \caption{
    (a) Schematic of a U(1)-GfPEPS projector $\ket{T_{\mathbf{r}}}$ together with the maximally entangled virtual bonds between neighboring sites.
    (b) The resulting fermionic PEPS on a square lattice by tiling the local tensors together.
  }
  \label{fig:gpeps}
\end{figure}

To define a U(1)-GfPEPS (see Fig.~\ref{fig:gpeps}), virtual fermions between every two neighboring sites form $M$ maximally entangled bonds,
\begin{equation}
    \vert I \rangle = \prod_{\mathbf{r}}\prod_{\alpha=1}^{M} (c^\dag_{\mathbf{r},r,\alpha} + c^\dag_{\mathbf{r}+\mathbf{x},l,\alpha})(c^\dag_{\mathbf{r},u,\alpha} + c^\dag_{\mathbf{r}+\mathbf{y},d,\alpha}) \vert0\rangle_\mathrm{v} ,
\label{eq:virtual-bond-state-1}
\end{equation}
where, for an $L\times L$ lattice with periodic or antiperiodic boundary conditions, virtual fermions have a fixed particle number $N_{\mathrm{v}} = 2ML^2$. 
$\vert0\rangle_{\mathrm{v}}$ is the vacuum of virtual fermions.
A fermionic PEPS is defined by $\vert \Psi \rangle = \langle I \vert T \rangle$~\cite{Kraus2010,Wahl2014}, where $\vert T \rangle$ is the PEPS projector
\begin{equation}
    \vert T \rangle = \prod_{\mathbf{r}} T_{\mathbf{r}} \vert 0 \rangle_{\mathrm{p},\mathrm{v}} .
\label{eq:GFPEPS-projector1}
\end{equation}
Here $\vert0\rangle_{\mathrm{p},\mathrm{v}}$ is the shared vacuum of physical and virtual fermions, and $T_{\mathbf{r}}$ creates a local state of physical and virtual fermions at site $\mathbf{r}$. 
For illustrating the construction, we shall focus on the translationally invariant case and consider the same $T_{\mathbf{r}}$ for all sites~\footnote{This can be straightforwardly generalized to the case of a larger unit cell, where different sites in the same unit cell have different $T_{\mathbf{r}}$.}.
The PEPS is hence fully characterized by the local state $T_{\mathbf{r}} \vert 0 \rangle_{\mathrm{p},\mathrm{v}}$.
Generally, $T_{\mathbf{r}}$ is parametrized as
\begin{align}
T &= \sum_{\{m_{\mu}\},\{n_{\nu,\alpha}\}} T^{\{m_{\mu}\}}_{\{n_{\nu,\alpha}\}} \nonumber \\
&\phantom{=} \times \left[ \prod_{\mu=1}^{P} 
(c_{\mu}^{\dag})^{m_{\mu}} \right]
\left[ \prod_{\nu=l,r,d,u} \prod_{\alpha=1}^{M} 
(c_{\nu,\alpha}^{\dag})^{n_{\nu,\alpha}} \right],
\label{eq:GFPEPS-projector2}
\end{align}
where, here and hereafter, the site index $\mathbf{r}$ is dropped when we refer to a local site.
$m_{\mu}$ ($n_{\nu,\alpha}$) is understood as the collection of occupation numbers of physical (virtual) modes.
The conserved fermion parity of $\vert \Psi \rangle$, known as the ``fermion superselection rule'', is imposed by requiring that $T^{\{m_{\mu}\}}_{\{n_{\nu,\alpha}\}}$ vanishes if $\sum_{\mu} m_{\mu} + \sum_{\nu,\alpha} n_{\nu,\alpha}$ is odd (or even).

For describing the ground state of fermionic systems with a fixed particle number, the $\mathbb{Z}_2$ parity conservation of the local tensor $T$ should be promoted to the U(1) particle-number conservation, by imposing that $T^{\{m_{\mu}\}}_{\{n_{\nu,\alpha}\}}$ is nonvanishing if and only if $\sum_{\mu} m_{\mu} + \sum_{\nu,\alpha} n_{\nu,\alpha} = Q$, where $Q$ is the total number of physical and virtual fermions at a single site. 
We henceforth restrict ourselves to free fermionic systems (i.e., ones described by quadratic fermionic Hamiltonians), and require the PEPS projector in Eq.~\eqref{eq:GFPEPS-projector1} to be a fermionic Gaussian state~\cite{Kraus2010}.
Thus, for PEPS describing free fermionic ground states with a fixed particle number, the PEPS projector reduces to a local Slater determinant created by
\begin{equation}
T = \prod_{q=1}^{Q} d^{\dag}_{q} \, ,
\label{eq:GFPEPS-projector3}
\end{equation}
where the orbitals $d^{\dag}_{q}$ are linear combinations of physical modes $c^{\dag}_{\mathbf{r},\mu}$ and virtual modes $c^{\dag}_{\mathbf{r},\nu,\alpha}$ at the same site.
The explicit form of $d^{\dag}_{q}$ will be specified in Sec.~\ref{subsec:correlation-matrix}.
For the U(1)-GfPEPS defined as $\vert \Psi \rangle = \langle I \vert T \rangle$, the number of physical fermions that remain after contracting the virtual modes is $N_{\mathrm{p}} = QL^2 - N_{\mathrm{v}} = (Q-2M)L^2$.
For a system of spin-1/2 fermions, the half-filling condition $N_{\mathrm{p}}=L^2$ is achieved by choosing $Q=2M+1$.

\subsection{Correlation matrix formulation}
\label{subsec:correlation-matrix}
As for fermionic Gaussian states, the virtual bond state $|I\rangle$ and PEPS projector $|T\rangle$ are characterized by their correlation matrices~\cite{Peschel2003,Bravyi2005}.
This provides an efficient computational tool for U(1)-GfPEPS.
Below we provide key results that are relevant for U(1)-GfPEPS and leave further details to Appendix~\ref{app:U1-Gaussian-formalism}.

Because of translational invariance, we switch to momentum space with $c^{\dag}_{\mathbf{r},\mu} = \frac{1}{L} \sum_{\mathbf{k}} c^{\dag}_{\mathbf{k},\mu} e^{-i\mathbf{k}\cdot \mathbf{r}}$ for physical modes ($\mu$ replaced by $\nu,\alpha$ for virtual modes).
$\mathbf{k}=(k_x,k_y)$ is a point in the first Brillouin zone (FBZ) and its allowed values depend on boundary conditions.
For instance, antiperiodic or periodic boundary conditions along the $x$-direction allow $k_x = \tfrac{2\pi}{L}(n_x+1/2)$ or $k_x = \tfrac{2\pi}{L}n_x$, respectively, with $n_x = 0, 1, \ldots, L-1$.

For the virtual bond state $|I\rangle$, we write its density operator as $\rho_{\mathrm{in}} = \ket{I}\bra{I}$ (input of U(1)-GfPEPS) and define its correlation matrix as
\begin{equation}
[\mathcal{C}_{\mathrm{in}}(\mathbf{k})]_{(\nu,\alpha),(\nu',\alpha')} = 2 \mathrm{tr}_{\mathrm{v}} (\rho_{\mathrm{in}}c^{\dag}_{\mathbf{k},\nu,\alpha} c_{\mathbf{k},\nu',\alpha'}) - \delta_{\nu,\nu'}\delta_{\alpha,\alpha'},
\end{equation}
where the trace $\mathrm{tr}_{\mathrm{v}}$ is with respect to virtual modes.
Such a correlation matrix is called a complex correlation matrix in Appendix~\ref{app:U1-Gaussian-formalism}.
To calculate this correlation matrix, one may express $|I\rangle$ in momentum space as
\begin{equation}
    \vert I \rangle = \prod_{\mathbf{k}}\prod_{\alpha=1}^{M} (c^\dag_{\mathbf{k},r,\alpha} + c^\dag_{\mathbf{k},l,\alpha}e^{-ik_x})(c^\dag_{\mathbf{k},u,\alpha} + c^\dag_{\mathbf{k},d,\alpha}e^{-ik_y}) \vert0\rangle_\mathrm{v} .
\label{eq:virtual-bond-state-2}
\end{equation}
The explicit form of the $4M \times 4M$ correlation matrix $\mathcal{C}_{\mathrm{in}}(\mathbf{k})$ is then obtained as:
\begin{equation}
 \mathcal{C}_{\mathrm{in}}(\mathbf{k}) = \left( \begin{matrix} 0 & e^{ik_x}\mathbbm{1}_{M} \\ e^{-ik_x}\mathbbm{1}_{M} & 0 \end{matrix} \right)  \oplus \left( \begin{matrix} 0 & e^{ik_y}\mathbbm{1}_{M} \\ e^{-ik_y}\mathbbm{1}_{M} & 0 \end{matrix} \right) ,
\label{eq:bond-correlation-matrix}
\end{equation}
where $\mathbbm{1}_{M}$ is an $M \times M$ identity matrix.

As the PEPS projector $|T\rangle$ assumes a translationally invariant onsite form [see Eq.~\eqref{eq:GFPEPS-projector1}], its correlation matrix is block diagonal in both real and momentum space, and all blocks are the same.
Thus, it is sufficient to parameterize this block by considering a single site $\mathbf{r}$ (or momentum $\mathbf{k}$):
\begin{equation}
 \mathcal{C}_{\mathrm{T}} = \left( \begin{matrix} A & B \\ B^{\dag} & D \end{matrix} \right).
\label{eq:projector-correlation-matrix-1}
\end{equation}
The submatrices encode two-point correlators between two physical modes ($P \times P$ matrix $A$), two virtual modes ($4M \times 4M$ matrix $D$), and one physical and one virtual mode ($P \times 4M$ matrix $B$):
\begin{align}
A_{\mu,\mu'} &= 2 \mathrm{tr}_{\mathrm{p},\mathrm{v}}(\rho_{\mathrm{T}}c^{\dag}_{\mathbf{r},\mu} c_{\mathbf{r},\mu'}) -\delta_{\mu,\mu'}, \nonumber \\
D_{(\nu,\alpha),(\nu',\alpha')} &= 2 \mathrm{tr}_{\mathrm{p},\mathrm{v}}(\rho_{\mathrm{T}}c^{\dag}_{\mathbf{r},\nu,\alpha} c_{\mathbf{r},\nu',\alpha'}) - \delta_{\nu,\nu'}\delta_{\alpha,\alpha'}, \nonumber \\
B_{\mu,(\nu',\alpha')} &= 2 \mathrm{tr}_{\mathrm{p},\mathrm{v}}(\rho_{\mathrm{T}}c^{\dag}_{\mathbf{r},\mu} c_{\mathbf{r},\nu',\alpha'}) ,
\label{eq:projector-correlation-matrix-2}
\end{align}
where $\rho_{\mathrm{T}}$ is the Gaussian density operator for $|T\rangle$ and $\mathrm{tr}_{\mathrm{p},\mathrm{v}}$ is with respect to both physical and virtual modes.
It is transparent that Eq.~\eqref{eq:projector-correlation-matrix-2} has the same form in momentum space (i.e., $\mathbf{r}$ replaced by $\mathbf{k}$).
Further important information utilizing the results in Appendix~\ref{app:U1-Gaussian-formalism} is as follows: As $|T\rangle$ is a pure state, $\mathcal{C}_{\mathrm{T}}$ is Hermitian and can be diagonalized as 
\begin{equation}
    U^{\dag}\mathcal{C}_{\mathrm{T}} U = \left( \begin{matrix} \mathbbm{1}_{Q} &  0 \\ 0 & -\mathbbm{1}_{P+4M-Q} \end{matrix} \right),
\label{eq:CT-diagonalization}
\end{equation}
where the identity block $\mathbbm{1}_{Q}$ corresponds to occupied single-particle orbitals $d^{\dag}_q$ [see Eq.~\eqref{eq:GFPEPS-projector3}]. 
Their explicit form is given by
\begin{equation}
d^{\dag}_{q}=\sum_{\mu=1}^{P} U^{\dagger}_{q,\mu} c^{\dag}_{\mu} + \sum_{\nu=l,r,d,u} \sum_{\alpha=1}^{M} U^{\dagger}_{q,(\nu,\alpha)} c^{\dag}_{\nu,\alpha}
\label{eq:occupied-orbitals}
\end{equation}
with $q=1,\ldots, Q$.

For the U(1)-GfPEPS $\vert \Psi \rangle = \langle I \vert T \rangle$, its Gaussian density operator $\rho_{\mathrm{out}}$ is obtained from $\rho_{\mathrm{out}}\propto \mathrm{tr}_{\mathrm{v}}(\rho_{\mathrm{T}}\rho_{\mathrm{in}})$ as the output.
The correlation matrix of $\rho_{\mathrm{out}}$ is block diagonal in momentum space and can be defined as
\begin{equation}
[\mathcal{C}_{\mathrm{out}}(\mathbf{k})]_{\mu,\mu'} = 2 \mathrm{tr}_{\mathrm{p}} (\rho_{\mathrm{out}}c^{\dag}_{\mathbf{k},\mu} c_{\mathbf{k},\mu'}) - \delta_{\mu,\mu'}.
\label{eq:GfPEPS-correlation-matrix-1}
\end{equation}
It is related to $\mathcal{C}_{\mathrm{in}}(\mathbf{k})$ and $\mathcal{C}_{\mathrm{T}}$ via
\begin{equation}
  \mathcal{C}_{\rm{out}}(\mathbf{k}) = A - B [D + \mathcal{C}_{\rm{in}}(\mathbf{k})]^{-1}B^{\dag},
\label{eq:GfPEPS-correlation-matrix-2}
\end{equation}
as shown in Appendix~\ref{app:U1-Gaussian-formalism}.
This expression is the main formal result of this paper.
 
Before moving on to numerical optimization, we comment on which systems the U(1)-GfPEPS Ansatz is suitable for. 
Eq.~\eqref{eq:virtual-bond-state-2} shows that each $\mathbf{k}$ point in the FBZ accommodates $2M$ virtual modes.
These virtual modes should be contracted with virtual modes in the U(1)-GfPEPS projector $\vert T\rangle$, where the latter has $Q$ physical and virtual modes at each $\mathbf{k}$ point.
Thus, after contracting the virtual modes, the U(1)-GfPEPS has $Q-2M$ physical modes for each $\mathbf{k}$ point.
This means that, for U(1)-GfPEPS, the number of occupied physical modes must be the \emph{same} at each $\mathbf{k}$ point.
While gapped band insulators and gapless semimetals (e.g., those with Dirac points) fulfill this requirement, the possibility of describing a Fermi surface is ruled out.
Although gapless fermions with a Fermi surface are known to violate the entanglement area law~\cite{Wolf2006,Gioev2006} and cannot be described by PEPS with a fixed bond dimension in the thermodynamic limit, our explicit construction nevertheless puts a stronger constraint on U(1)-GfPEPS: If translational symmetry is preserved, U(1)-GfPEPS cannot have a Fermi surface even on finite-size systems. 

\subsection{Optimization}
Consider a quadratic Hamiltonian of fermions
\begin{equation}
H = \sum_{\mathbf{k}} \sum_{\mu,\mu'=1}^{P} c^{\dag}_{\mathbf{k},\mu} [\mathcal{H}(\mathbf{k})]_{\mu,\mu'} c_{\mathbf{k},\mu'} ,
\label{eq:target-Hamiltonian}
\end{equation}
where $\mathcal{H}(\mathbf{k})$ is the single-particle Hamiltonian matrix. 
We use the U(1)-GfPEPS as a variational ansatz to approximate its ground state.
We note that the U(1)-GfPEPS has $Q-2M$ occupied physical modes at each $\mathbf{k}$ point, so it will approximate the Fermi sea ground state of Eq.~\eqref{eq:target-Hamiltonian} with $Q-2M$ occupied bands, implying a filling factor $(Q-2M)/P$.
The variational energy of the U(1)-GfPEPS with correlation matrix $\mathcal{C}_{\mathrm{out}}$ [see Eq.~\eqref{eq:GfPEPS-correlation-matrix-1}] is given by
\begin{equation}
E = \frac{1}{2} \sum_{\mathbf{k}}{\rm{Tr}}[(\mathcal{C}_{\rm{out}}(\mathbf{k}) + \mathbbm{1}_P )\mathcal{H}(\mathbf{k})^T],
  \label{eq:Energy}
\end{equation}
where $\mathrm{Tr}$ is the usual matrix trace.
The variational space is the correlation matrix $\mathcal{C}_{\mathrm{T}}$ for the U(1)-GfPEPS projector~\eqref{eq:projector-correlation-matrix-1}, which relates to $\mathcal{C}_{\mathrm{out}}(\mathbf{k})$ via Eq.~\eqref{eq:GfPEPS-correlation-matrix-2} [$\mathcal{C}_{\mathrm{in}}(\mathbf{k})$ is fixed; see Eq.~\eqref{eq:bond-correlation-matrix}].

For the energy minimization, we observe that the unitary matrix $U$ in Eq.~\eqref{eq:CT-diagonalization} can be parameterized as $U = (W, W_{\perp})$, with $W$ corresponding to the occupied modes and $W_{\perp}$, the orthogonal complement of $W$, to the unoccupied ones.
By that, we can express $\mathcal{C}_{\mathrm{T}}$ in terms of $W$,
\begin{equation}
\mathcal{C}_{\mathrm{T}} = WW^{\dag} - W^{\phantom{}}_{\perp}W^{\dag}_{\perp} = 2WW^{\dag}-\mathbbm{1}_{P+4M}.
\end{equation}
Combining Eqs.~\eqref{eq:projector-correlation-matrix-1}, \eqref{eq:GfPEPS-correlation-matrix-2} and \eqref{eq:Energy}, our task boils down to numerically optimize $W$ to minimize the ground-state energy in Eq.~\eqref{eq:Energy} under the isometry constraint $W^{\dag}W=\mathbbm{1}_{Q}$.

We obtain the optimal $W$ by gradient based optimization schemes developed in Refs.~\cite{Edelman1998,VOORHIS2002,Abrudan2009,Zhu2016,Hauru2021}.
First, we compute the numerical gradient $g^{*}=\frac{\partial E}{\partial W}$, which can be evaluated by finite difference or auto-differentiation.
The gradients with respect to the unoccupied modes are always zero as they do not participate in the energetics.

Second, we project $g$ onto the tangent space of $U = (W, W_{\perp})$, which yields 
\begin{equation}
  G = (g-Wg^{\dagger}W, -Wg^{\dagger}W_{\perp}).
\end{equation}
Note that the equation defining tangent vectors $\Delta$ of $U$ can be obtained by differentiating $UU^{\dagger} = \mathbbm{1}$, which gives $\Delta U^{\dagger} + U\Delta^{\dagger}=0$ (i.e., $\Delta U^{\dagger}$ is skew-symmetric), and we can verify that $G$ indeed satisfies such a constraint.

Next, we minimize the energy along the geodesic defined by $G$, i.e., $E(\alpha)$, with $W(\alpha) = e^{-\alpha Q_{G}}W$, where 
\begin{equation}
Q_{G} = GU^{\dagger} = gW^{\dagger}-Wg^{\dagger}.
\end{equation}
The isometry $W$ is then updated according to the optimal value of $\alpha$ via the Wolfe line search~\cite{Nocedal2006}.
This procedure is repeated until the norm of the gradient is sufficiently small.
To accelerate the convergence of such gradient descent minimization, one can modify the line search direction by combining the current gradient with the previous ones;
commonly used methods include the nonlinear conjugate gradient~\cite{Edelman1998,Abrudan2009}, the Limited-memory Broyden–Fletcher–Goldfarb–Shanno~\cite{Hauru2021}, and the direct inversion in the iterative subspace~\cite{Csaszar1984}.
To reduce the numerical noise, one can anti-symmetrize $Q_{G}$ manually at the end, after adding up the gradients. 
All methods improve the convergence rate comparing to gradient descent.
In this work, we adopt the nonlinear conjugate gradient algorithm, and to reduce the numerical noise, we manually anti-symmetrize $Q_{G}$ at the end, after adding up the gradients.

Once the optimal $\mathcal{C}_{\mathrm{T}}$ and $\mathcal{C}_{\mathrm{out}}$ have been obtained, it is also possible to compare the exact band dispersions obtained by diagonalizing $\mathcal{H}(\mathbf{k})$ with the variational ones obtained from U(1)-GfPEPS. 
One can diagonalize $\mathcal{C}_{\mathrm{out}}(\mathbf{k})$ to obtain
\begin{equation}
V(\mathbf{k})^{\dagger} \mathcal{C}_{\mathrm{out}}(\mathbf{k}) V(\mathbf{k}) = \left( \begin{matrix} \mathbbm{1}_{Q-2M} &  0 \\ 0 & -\mathbbm{1}_{P-Q+2M} \end{matrix} \right).
\end{equation}
Then, the occupied physical orbitals are given by $f^{\dag}_{\mathbf{k},q}=\sum_{\mu=1}^{P} V(\mathbf{k})^{\dagger}_{q,\mu} c^{\dag}_{\mathbf{k},\mu}$ with $q=1,\ldots,Q-2M$.
The single-particle Hamiltonian $\mathcal{H}(\mathbf{k})$ is then projected into this one-particle-occupied subspace by defining  
\begin{equation}
[\mathcal{H}(\mathbf{k})]_{q,q'} =[V(\mathbf{k})^{\dagger}\mathcal{H}(\mathbf{k})V(\mathbf{k})]_{q,q'}
\end{equation}
with $q,q'=1,\ldots,Q-2M$.
Its eigenvalues give the variational dispersions for the \emph{filled} bands.

\subsection{Gutzwiller projection and tensor network contraction}
\label{subsec:Gutzwiller-Projection}
The Gutzwiller projection is implemented by a product of local operators.
For simplicity, we illustrate its implementation for spin-1/2 fermions at each site ($P=2$).
The full Gutzwiller projection is defined by $P_{\mathrm{G}} = \prod_{\mathbf{r}}(n_{\mathbf{r},\uparrow}-n_{\mathbf{r},\downarrow})^2$ with $n_{\mathbf{r},\mu} = c^{\dag}_{\mathbf{r},\mu} c_{\mathbf{r},\mu}$ ($\mu = \uparrow,\downarrow$).
$P_{\mathrm{G}}$ deletes empty and doubly occupied states and keeps two singly occupied states $|\mu\rangle = c^{\dag}_{\mu}|0\rangle$ that are identified as spin-1/2 degrees of freedom.

Once the U(1)-GfPEPS projector $\prod_{\mathbf{r}} T_{\mathbf{r}} \vert 0 \rangle_{\mathrm{p},\mathrm{v}}$ is obtained, the Gutzwiller projection results in a (fermionic) PEPS with projector $\prod_{\mathbf{r}} (n_{\mathbf{r},\uparrow}-n_{\mathbf{r},\downarrow})^2 T_{\mathbf{r}} \vert 0 \rangle_{\mathrm{p},\mathrm{v}}$, and the virtual bond state $\vert I \rangle$ is unchanged.
Utilizing this idea, it becomes possible to convert a Gutzwiller-projected Fermi sea state into PEPS, where the unprojected Fermi sea is approximated by optimizing U(1)-GfPEPS with respect to some quadratic Hamiltonians of fermions.

The remaining task is to derive the explicit tensor form of the U(1)-GfPEPS projector.
If we write the occupied orbitals in Eq.~\eqref{eq:occupied-orbitals} in a more compact form $d^{\dag}_{q} = \sum_{\zeta=1}^{P+4M} U_{q,\zeta}^{*} c^{\dag}_{\zeta}$ with $\zeta$ enumerating all physical and virtual modes, the U(1)-GfPEPS local projector in Eq.~\eqref{eq:GFPEPS-projector3} can be expressed in a Slater determinant form
\begin{equation}
T = \sum_{\zeta_1 < \cdots < \zeta_Q} \det(U^{\dagger}_{(1,\ldots,Q),(\zeta_1,\ldots,\zeta_Q)}) c^{\dag}_{\zeta_1} \cdots c^{\dag}_{\zeta_{Q}},
\label{eq:Slater-determinant}
\end{equation}
where local tensor coefficients [see Eq.~\eqref{eq:GFPEPS-projector2}] can be read out from the determinants.
Gutzwiller projection simply removes some configurations in Eq.~\eqref{eq:Slater-determinant}.
Other local operators can be applied in a similar way.

\begin{figure}[!htb]
  \includegraphics[width=0.9\linewidth,trim = 0.0in 0.0in 0.0in 0.0in,clip=true]{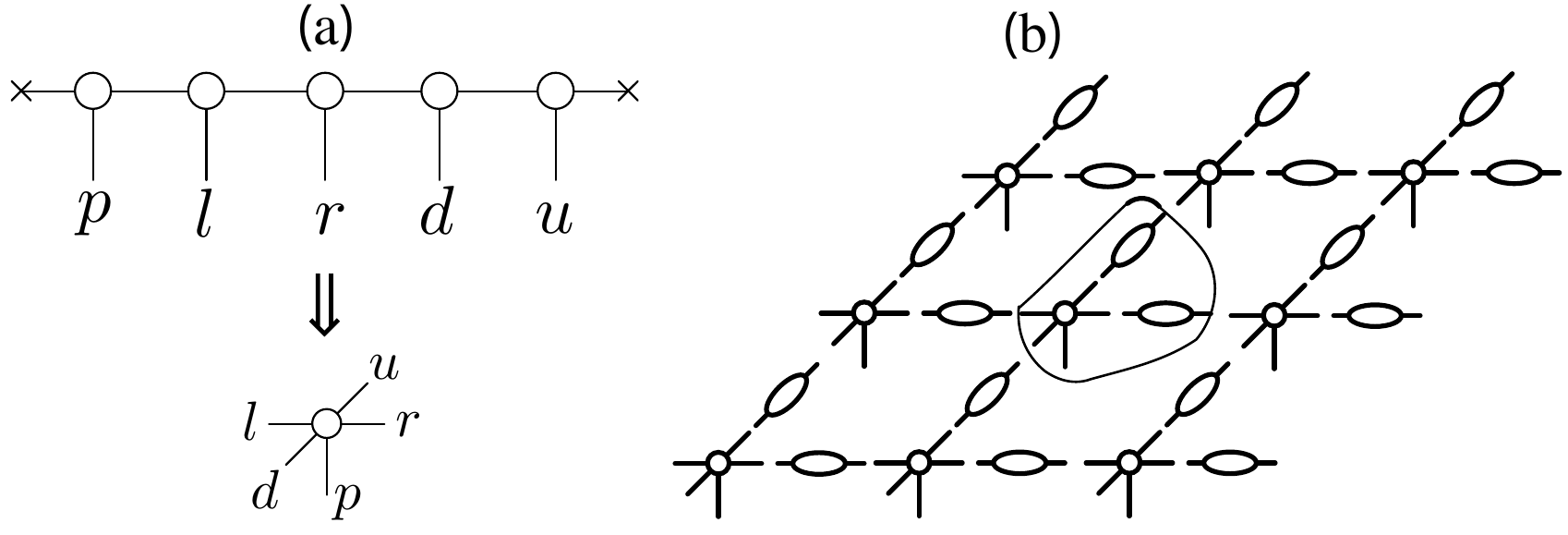}
  \caption{Schematics of (a) converting the MPS form of $T_{\mathbf{r}}$ to a PEPS local tensor and
  (b) contracting $T_{\mathbf{r}}$ with entangled bond states (in oval shapes) to obtain a PEPS represented by a single local tensor.
  }
  \label{fig:gPEPS2}
\end{figure}

Alternatively, one can also construct the U(1)-GfPEPS projector via the MPO--MPS approach~\cite{Wu2020}. 
This is most convenient when working with tensor network libraries supporting U(1) or non-Abelian symmetries.
For the local projector 
$T \vert 0 \rangle_{\mathrm{p},\mathrm{v}} = \prod_{q=1}^{Q} d^{\dag}_{q} \vert 0 \rangle_{\mathrm{p},\mathrm{v}}$, the vacuum $\vert 0 \rangle_{\mathrm{p},\mathrm{v}}$ is treated as an MPS with bond dimension $D=1$.
Each occupied orbital $d^{\dag}_{q}$ is then represented as an MPO with bond dimension $D=2$ (see Refs.~\cite{Wu2020,Chen2021}).
After applying all $Q$ MPOs for occupied orbitals, $T\vert 0 \rangle_{\mathrm{p},\mathrm{v}}$ is represented as an MPS with bond dimension $D=2^Q$. The local tensor in Eq.~\eqref{eq:GFPEPS-projector2} is obtained by contracting all virtual indices of this MPS [see Fig.~\ref{fig:gPEPS2}(a)].
The advantage of the MPO--MPS approach is that the tensor entries of $T\vert 0 \rangle_{\mathrm{p},\mathrm{v}}$ as well as the corresponding symmetry structure, including the quantum numbers of the symmetric tensors and the corresponding Clebsch–Gordan coefficients, can be automatically generated.

After the Gutzwiller projection, it is practical to contract the virtual bonds in Eq.~\eqref{eq:virtual-bond-state-1} into the PEPS local tensors [see Fig.~\ref{fig:gPEPS2}(b)].
As the optimization of U(1)-GfPEPS is very efficient and the system size that can be reached is quite large, we can tile up the resulting Gutzwiller-projected U(1)-GfPEPS tensor to approximate the state on an infinite lattice.
Such infinite PEPS involves a single tensor at each site and is ready for computing physical quantities with fermionic tensor network contraction algorithms~\cite{Corboz2010}.
For this work, we adopt the corner transfer matrix renormalization group (CTMRG) method~\cite{Nishino1996,Orus2009} to perform tensor network contractions. 
To achieve higher accuracy and reduce computational cost in CTMRG calculations, we impose both the U(1) particle-number and the SU$(2)$ spin symmetry provided by the QSpace libary~\cite{Weichselbaum2012,Weichselbaum2020}.

\section{Results}
\label{sec:results}

\subsection{Dirac fermion models on square and kagome lattices}
\label{subsec:Dirac-FS}

\begin{figure}[!htb]
  \includegraphics[width=0.9\linewidth,trim = 0.0in 0.0in 0.0in 0.0in,clip=true]{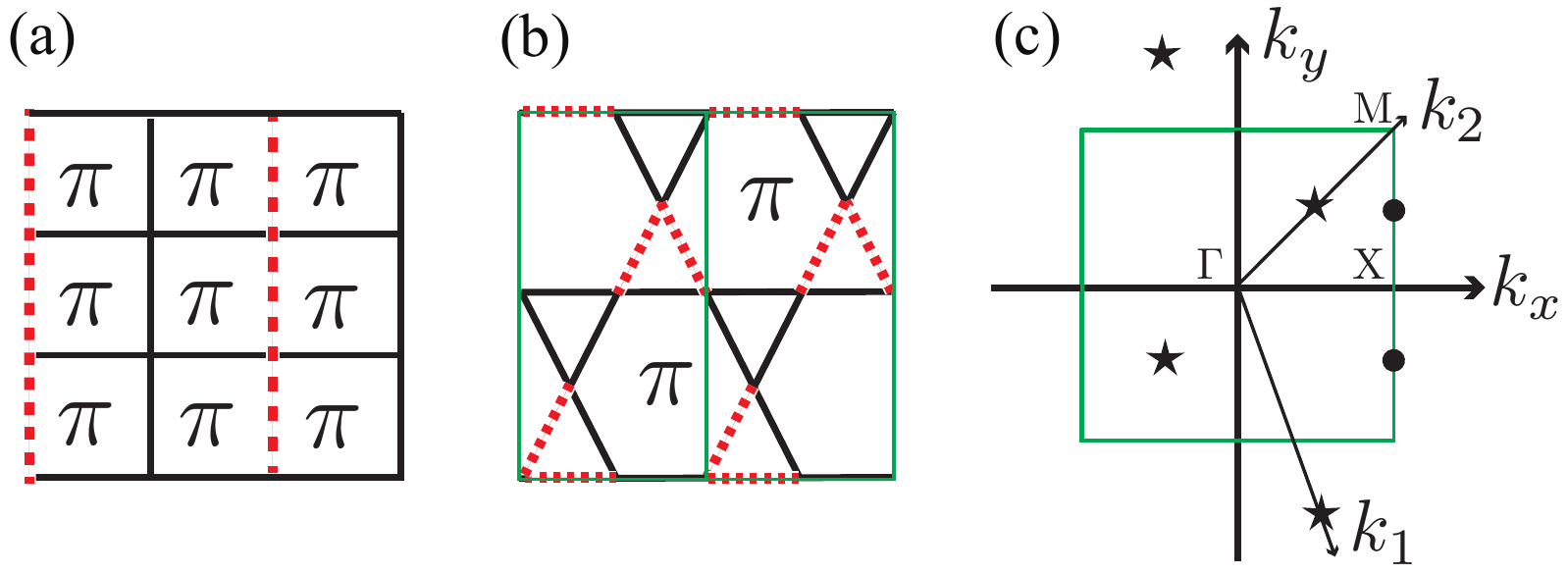}
  \caption{Schematics of (a) the $\pi$-flux model on the square lattice and (b) the $[0,\pi]$-flux model on the kagome lattice.
  The solid (red dashed) lines are the bonds with hopping $t=1$ ($t=-1$).
  (c) The first Brillouin zone (in green) of the effective square lattices for (a) and (b) 
  with $\Gamma = (0,0)$, $X=(\pi,0)$ and $M=(\pi,\pi)$.
  The black dots denote two Dirac nodes at $(\pi,\pm \pi/2)$ for the $\pi$-flux model, and the black stars at $(\pi/2, -3\pi/2)$ and $(\pi/2, \pi/2)$ for the $[0,\pi]$-flux model along $k_1$ and $k_2$ directions, respectively.
  }
  \label{fig:lattice}
\end{figure}

As benchmark examples, we use U(1)-GfPEPS to approximate the Fermi sea ground states of two spinless fermion models with a Dirac spectrum: the $\pi$-flux model on the square lattice~\cite{Affleck1988} and the $[0,\pi]$-flux model on the kagome lattice~\cite{Hastings2000,Ran2007}.
Both models have nearest-neighbor hoppings with the Hamiltonian
\begin{equation}
 H = \sum_{\langle \mathbf{r},\mathbf{r}' \rangle}t_{\mathbf{r},\mathbf{r}'}c^{\dag}_\mathbf{r}c_{\mathbf{r}'},
\label{eq:Dirac-Hamiltonians}
\end{equation}
where the square-lattice model has $\pi$-flux within each plaquette, and the kagome model has zero flux through each triangle and $\pi$-flux through each hexagon.
The hoppings realizing these flux choices are depicted in Figs.~\ref{fig:lattice}(a) and (b).

The $\pi$-flux square-lattice ($[0,\pi]$-flux kagome) model has a two-site (six-site) unit cell.
We group all sites in the same unit cell together and treat them as a single site in an effective square lattice.
This allows us to use a translationally invariant U(1)-GfPEPS ansatz with $P=2$ ($P=6$) for the $\pi$-flux square-lattice ($[0,\pi]$-flux kagome) model.
At half filling, both models have two Dirac nodes in the FBZ, as shown in Fig.~\ref{fig:lattice}(c).
For the numerical optimization, the effective square lattice has size $L\times L$ and the boundary condition (periodic or antiperiodic) is adjusted such that exact zero-energy modes at the Dirac nodes are avoided to ensure a unique ground state.
The optimal U(1)-GfPEPS is determined numerically for each fixed number of virtual mode $M$, when the averaged norm of its energy gradient with respect to the Hamiltonian in Eq.~\eqref{eq:Dirac-Hamiltonians} is smaller than $10^{-6}$.

\begin{figure}[!htb]
  \includegraphics[width=1.0\linewidth,trim = 0in 0in 0in 0in,clip=true]{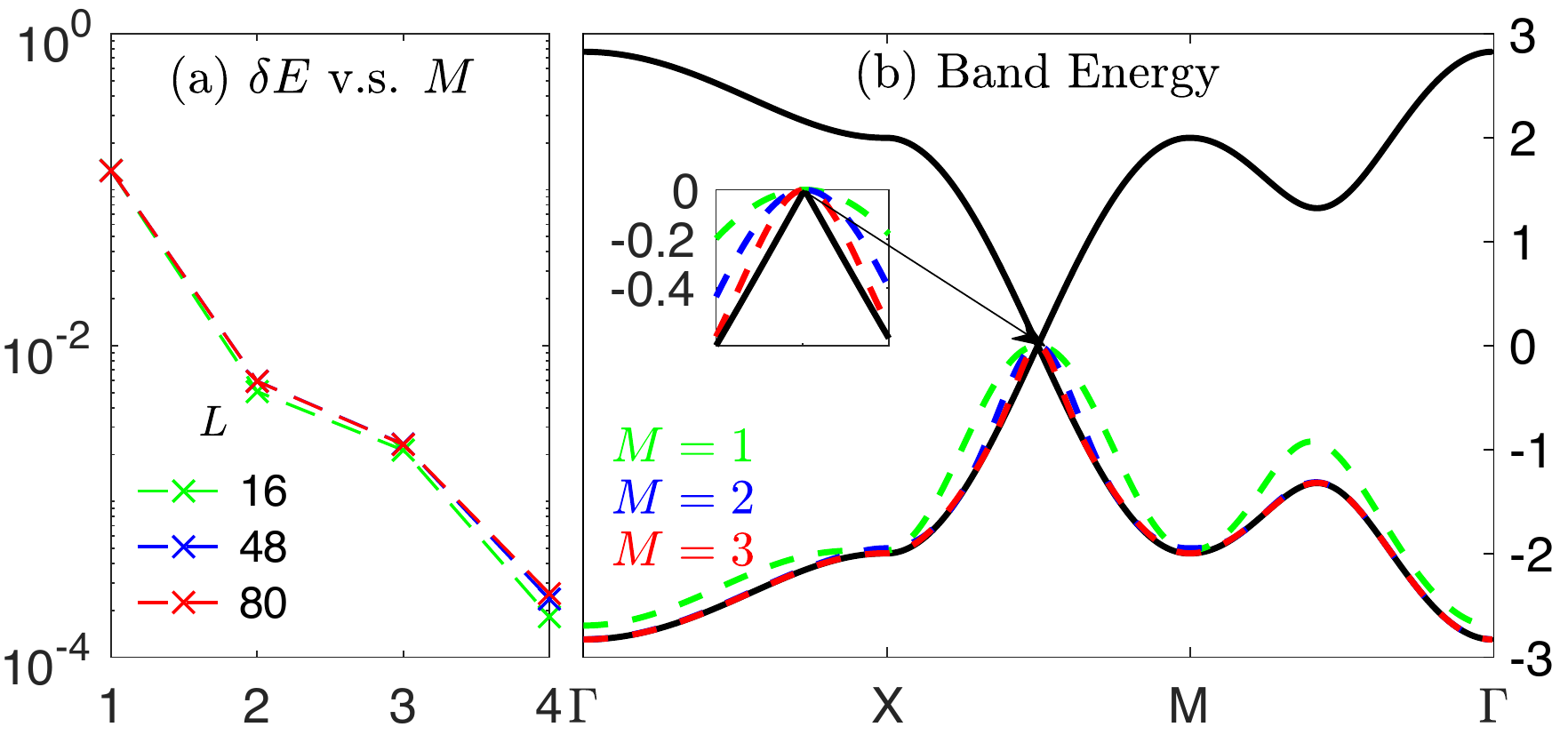}
  \caption{Results of optimized U(1)-GfPEPS for the $\pi$-flux state on the square lattice.
    (a) Relative error in the energy density of U(1)-GfPEPS versus the number of virtual modes, $M$.
    (b) Plots of the exact band structure (solid lines) and the variationally obtained occupied band at half filling (dashed lines).
  }
  \label{fig:SQ1}
\end{figure}

For the $\pi$-flux square-lattice model, we observe that the relative error in the ground-state energy density $\delta E$ decreases exponentially when increasing the number of virtual modes $M$ [see Fig.~\ref{fig:SQ1}(a)].
Furthermore, the finite-size effect in the energy density error appears to be small.
As shown in Fig.~\ref{fig:SQ1}(b), the U(1)-GfPEPS with $M=2$ (bond dimension $D=4$), which is variationally optimized on a $80\times 80$ lattice, reproduces the band dispersion in the thermodynamic limit very well.

\begin{figure}[!htb]
  \includegraphics[width=1.0\linewidth,trim = 0in 0in 0in 0in,clip=true]{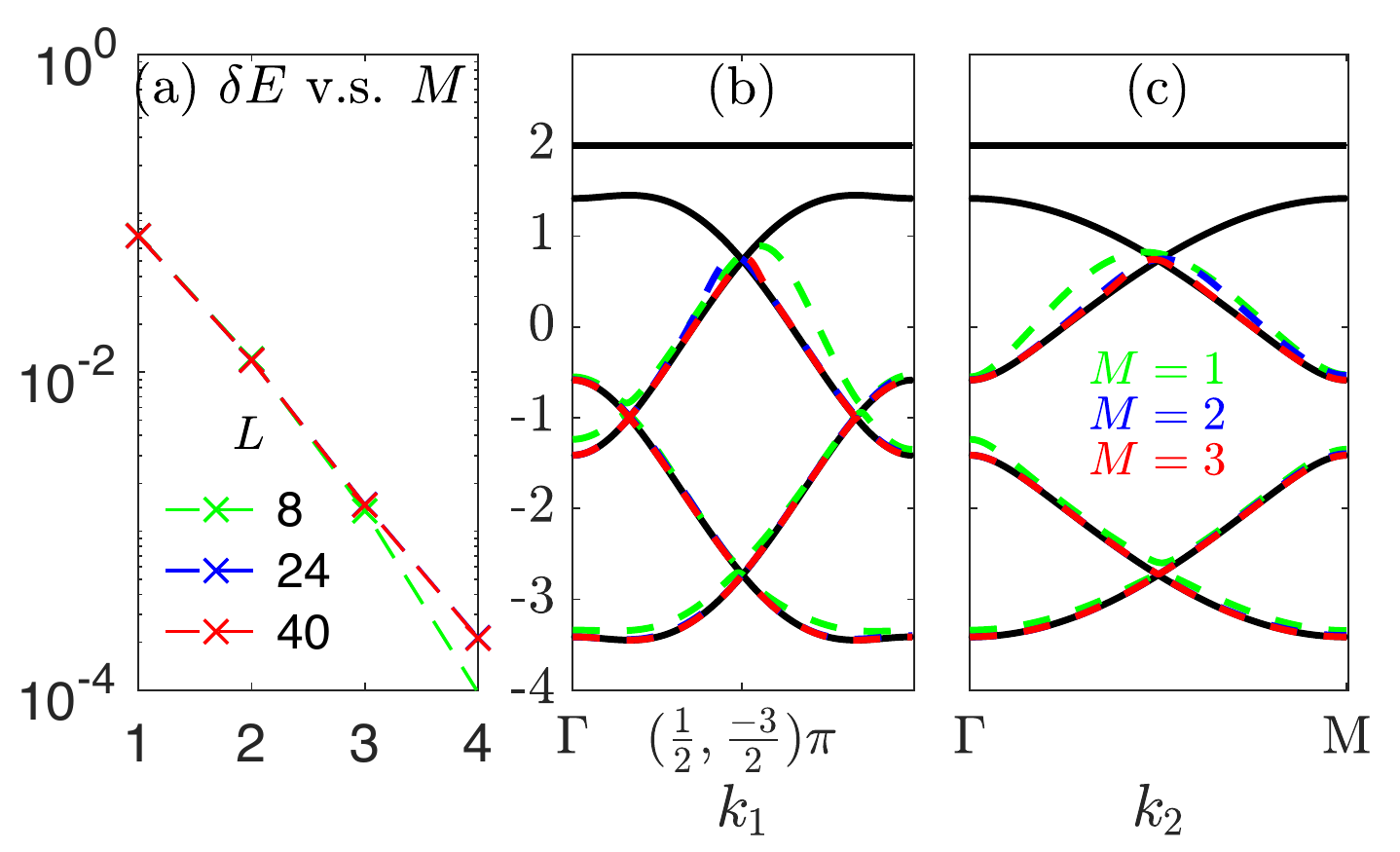}
  \caption{Results of optimized U(1)-GfPEPS for the $[0,\pi]$-flux model on the kagome lattice.
    (a) Relative error in the energy density of U(1)-GfPEPS versus the number of virtual modes, $M$.
    (b), (c) Plots of the exact band structure (solid lines) and the variationally obtained lower occupied bands (dashed lines), as functions of $k_1$ and $k_2$.
  }
  \label{fig:KG1}
\end{figure}

For the $[0,\pi]$-flux kagome model, the relative error of the ground-state energy density $\delta E$ in Fig.~\ref{fig:KG1}(a) follows the same trend as of the $\pi$-flux square-lattice model.
At half filling, the low-energy physics is dictated by two Dirac nodes [Fig.~\ref{fig:KG1}(b)].
The band dispersions along $k_1$ and $k_2$ directions (cutting two Dirac nodes) are plotted in Figs.~\ref{fig:KG1}(b) and (c).
With that, we examine the results due to the U(1)-GfPEPS approximation at small $M$. 
We again find a good agreement between the variational results with $M=2$ and the exact solution in the thermodynamic limit.

\subsection{U(1)-Dirac spin liquids on square and kagome lattices}

The optimized U(1)-GfPEPS for Dirac Fermi sea states in Sec.~\ref{subsec:Dirac-FS} are then used to build PEPS representing U(1)-Dirac spin liquids.
To this end, we attach a spin index $\sigma = \uparrow,\downarrow$ to the physical modes in Eq.~\eqref{eq:Dirac-Hamiltonians} and interpret them as fermionic partons for a spin-1/2 system.
The spin-1/2 operators are written as $\mathbf{S}(\mathbf{r}) = \frac{1}{2}\sum_{\sigma \sigma'} c^{\dag}_{\mathbf{r}\sigma} \boldsymbol{\tau}_{\sigma \sigma'}c_{\mathbf{r}\sigma'}$, where $\boldsymbol{\tau}$ are Pauli matrices.
The single-occupancy constraint $\sum_{\sigma} c^{\dag}_{\mathbf{r}\sigma} c_{\mathbf{r}\sigma}=1$ ensures the physical spin-1/2 Hilbert space and is imposed by the full Gutzwiller projection.

Starting from a U(1)-GfPEPS $\ket{\Psi}$ for spinless fermions, we just need two copies of it (with different spins) and apply the Gutzwiller projection to obtain a PEPS for spin-1/2 system, i.e., $\ket{\Phi} = P_{\mathrm{G}} \ket{\Psi_{\uparrow}}\otimes\ket{\Psi_{\downarrow}}$.
For $\ket{\Psi}$ with virtual bonds and projector defined in Eqs.~\eqref{eq:virtual-bond-state-1} and \eqref{eq:GFPEPS-projector3}, $\ket{\Psi_{\uparrow}}\otimes\ket{\Psi_{\downarrow}}$ is obtained by attaching a spin index $\sigma$ to both virtual and physical modes, e.g., the projector with $T = \prod_{q=1}^{Q}\prod_{\sigma=\uparrow,\downarrow} d^{\dag}_{q,\sigma}$ (similar for the virtual bonds).
If $\ket{\Psi}$ has bond dimension $D=2^M$, the Gutzwiller-projected PEPS $\ket{\Phi}$ has bond dimension $D = 4^M$. The method for determining the local tensor of $\ket{\Phi}$ is described in Sec.~\ref{subsec:Gutzwiller-Projection}.

As the U(1)-GfPEPSs obtained in Sec.~\ref{subsec:Dirac-FS} represent Dirac Fermi sea states, it is possible to obtain U(1)-Dirac spin liquids after the Gutzwiller projection~\cite{Ran2007,Hermele2008}.
The field theory governing the large-distance behavior of U(1)-Dirac spin liquids is quantum electrodynamics in 2+1 dimensions (QED$_3$), with $N_{\mathrm{f}}$-flavor Dirac fermions coupled to a U(1) gauge field.
The calculation of critical exponents in QED$_3$ is, however, very challenging, especially when the fermion flavor $N_{\mathrm{f}}$ is not large~\cite{Hermele2004}. 
As our setups in Sec.~\ref{subsec:Dirac-FS} have two Dirac nodes, the Gutzwiller-projected U(1)-GfPEPSs should be relevant to QED$_3$ with $N_{\mathrm{f}}=4$. It is thus an interesting task to compute their critical exponents with PEPS techniques in the thermodynamic limit. 

We focus in this work on the staggered spin-spin correlation function $C(r) = (-1)^{r}\braket{\mathbf{S}(0)\cdot \mathbf{S}(\mathbf{r})}$, where two spins, with distance $r=\vert \mathbf{r} \vert $, are placed on the same row of the effective square lattice.
Due to the large computational cost, we have only performed calculations using Gutzwiller-projected U(1)-GfPEPSs with $D=4$ and $16$ ($M=1$ and $2$).
For a given $D$, we compute the environment of PEPS via CTMRG method with a fixed number of symmetry multiplets $\chi^{*}$, which roughly corresponds to a typical bond dimension of $\chi=2\chi^{*}$ if symmetries are not used.
The CTMRG environment constitutes the bulk part of the infinite lattice, and its accuracy can be examined by varying $\chi^{*}$.

\begin{figure}[!htb]
  \includegraphics[width=1.0\linewidth,trim = 0in 0in 0in 0in,clip=true]{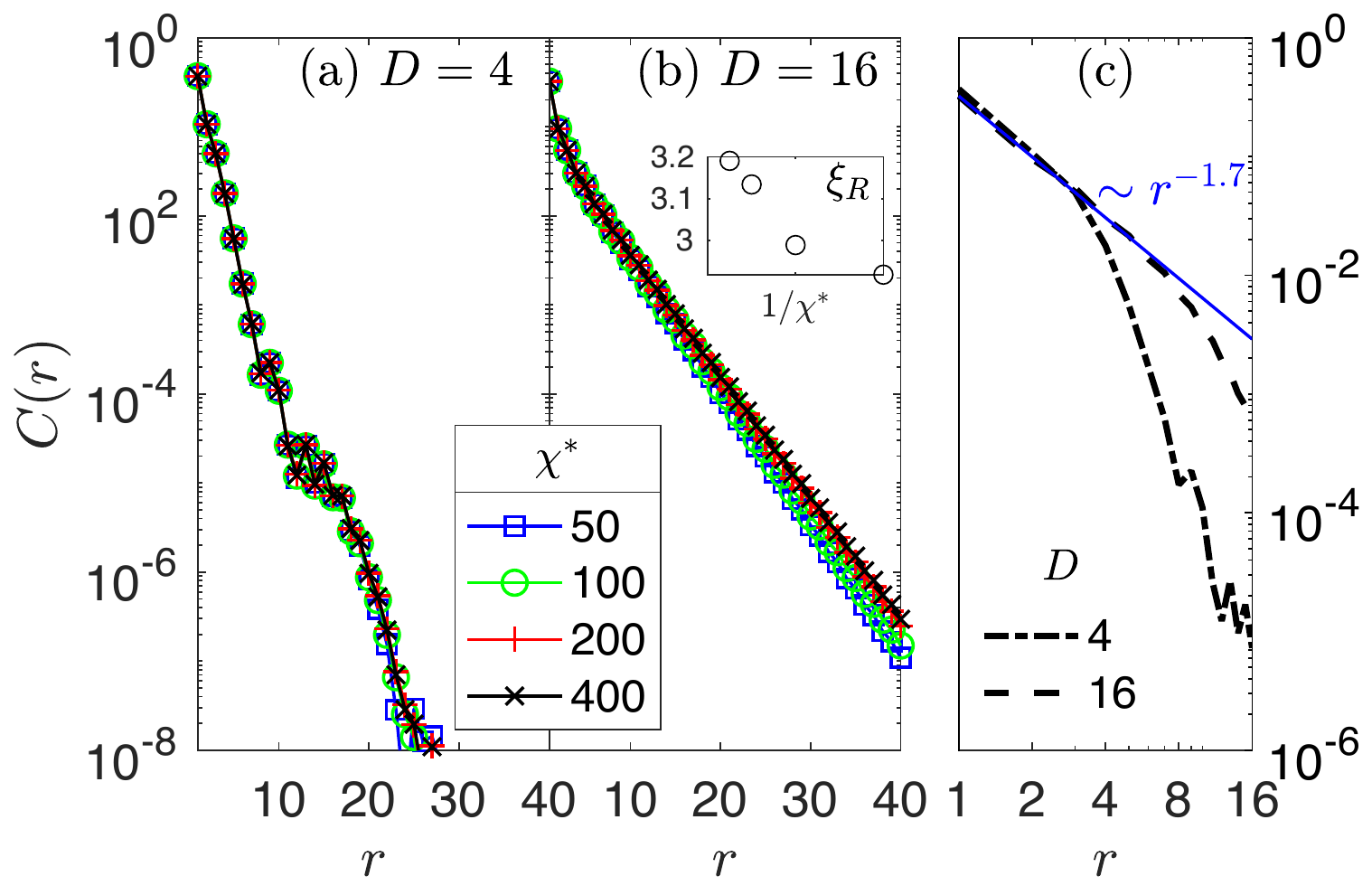}
  \caption{Staggered spin correlations for Gutzwiller-projected U(1)-GfPEPS from the $\pi$-flux state on the square lattice.
    Semilogarithmic plots for (a) $D=4$ and (b) $D = 16$ with different environmental bond dimensions $\chi^{*}$.
    (c) Log-log plots with $\chi^{*}=400$, and the blue solid line shows a powerlaw decay with an exponent $\eta = 1.7$.
  }
  \label{fig:SQ2}
\end{figure}

For the PEPS representing Gutzwiller projected $\pi$-flux state, the results are plotted in Figs.~\ref{fig:SQ1}(a) and (b).
For $D=4$, $C(r)$ has a fast exponential decay, which is almost unchanged when varying $\chi^{*}$.
However, such exponential decay gets slowed down as we increase the bond dimension to $D=16$.
We also observe an increase of the correlation length $\xi_R$ at large distance ($C(r) \sim e^{-r/\xi_R}$) by increasing $\chi^{*}$.
Overall, our results suggest that the spin gap imposed by the finite bond dimensions ($D$ and $\chi^{*}$) can be further reduced.
However, at this stage, we cannot predict to which value of $D$ one may achieve an algebraic decay at large distance.
Turning to the short distance regime [Fig.~\ref{fig:SQ2}(c)], we observe a buildup of a powerlaw decay $C(r) \sim r^{-\eta}$ with exponent $\eta \approx 1.7$. 
This is in rough agreement with previous Monte Carlo estimates ($\eta \approx 1.6$~\cite{Ivanov1999} and $\eta \approx 2$~\cite{Ferrari2021}) on finite-size clusters, but smaller than the extrapolation of the large-$N_{\mathrm{f}}$ result $\eta = 4-64/(3\pi^2 N_{\mathrm{f}}) + \mathcal{O}(1/N_{\mathrm{f}}^2)$~\cite{Rantner2002} to $N_{\mathrm{f}}=4$, which gives $\eta \approx 3.46 $.

\begin{figure}[!htb]
  \includegraphics[width=1.0\linewidth,trim = 0in 0in 0in 0in,clip=true]{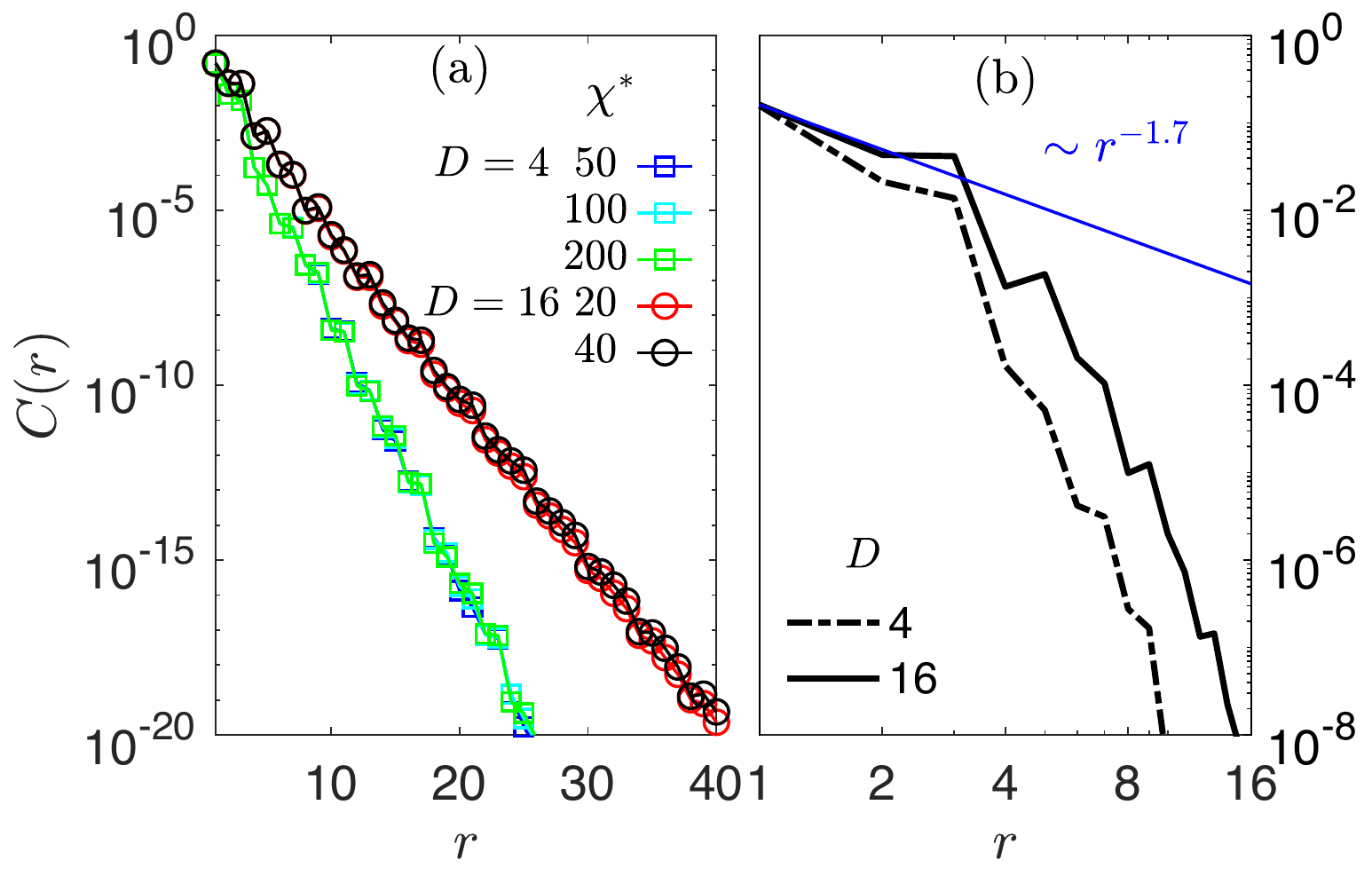}
  \caption{Staggered spin correlations for Gutzwiller-projected U(1)-GfPEPS from the $[0,\pi]$-flux state on the kagome lattice.
  (a) Semilogarithmic plots for $D=4$ and $D=16$ with different environmental bond dimension $\chi^{*}$.
  (b) Log-log plots with the largest possible $\chi^{*}$. The blue solid line showing the powerlaw decay with exponent $\eta = 1.7$ is a guide to the eye.
  }
  \label{fig:KG2}
\end{figure}

For the kagome-lattice case, the calculation with the Gutzwiller-projected U(1)-GfPEPS is very challenging, since the physical index of each PEPS local tensor contains six spin-1/2's (physical dimension $d=32$). This makes it difficult to contract double layer tensors in CTMRG.
Therefore, for $D=16$, we only report results with small environmental bond dimensions $\chi^{*}=20$ and $40$.
Nevertheless, in Fig.~\ref{fig:KG2}(a), one can still observe an increase in the correlation length when going from $D=4$ to $16$.
This entails a rather severe finite $D$ effect, similar to the square-lattice case. 
From the plot in log-log scale [Fig.~\ref{fig:KG2}(b)], we see a quick deviation from the powerlaw behavior. Thus, for the Gutzwiller projected $[0,\pi]$-flux state, reliable conclusions cannot be made from these results.
This issue, instead, should be further investigated with even larger bond dimensions, which is beyond our current computational capability.

\section{Summary and outlook}
\label{sec:summary}

To summarize, we have put forward a formalism for constructing particle-number-conserving Gaussian fermionic projected entangled pair states. These states are suitable for describing the ground states of gapped band insulators and gapless fermions with band touching points, but incapable of describing gapless fermions with a Fermi surface.
We further develop a systematic method using these states as variational Ans\"atze for approximating the Fermi sea ground states of free fermionic Hamiltonians.
Benchmark calculations on the $\pi$-flux square-lattice model and the $[0,\pi]$-flux kagome-lattice model have shown excellent results.
The implementation of additional Gutzwiller projection on top of these variationally obtained U(1)-GfPEPS provides PEPS representation of U(1)-Dirac spin liquid states for spin-1/2 systems.
Using the CTMRG method to calculate spin-spin correlation functions in the thermodynamic limit, we have obtained a critical exponent $\eta \approx 1.7$ from the Gutzwiller-projected U(1)-GfPEPS representing $\pi$-flux U(1)-Dirac spin liquid state on the square lattice.

Computationally, the optimization of the U(1)-GfPEPS using correlation matrix is efficient, as the number of parameters scales linearly with respect to the number of virtual modes $M$. 
The size of real-space PEPS tensor, on the other hand, grows exponentially when increasing $M$.
This turns out to be the bottleneck for constructing the Gutzwiller-projected U(1)-GfPEPS with larger bond dimensions.

For future works, one interesting direction is to use our method to test the quality of Gutzwiller projected wave functions for challenging strongly correlated systems, such as the kagome Heisenberg antiferromagnet and the $t$--$J$ model.
It is also a promising direction to use them as initial Ans\"atze to improve the performance of PEPS variational algorithms.

\section*{Acknowledgments}
We thank Nick Bultinck, Meng Cheng, and Jutho Haegeman for helpful discussions. H.-H.T. is grateful to Lei Wang and Qi Yang for collaborations and stimulating discussions on a closely related topic. The numerical simulations in this work are based on the QSpace tensor library~\cite{Weichselbaum2012,Weichselbaum2020}. J.-W.L. and J.v.D. are supported by the Deutsche Forschungsgemeinschaft under Germany's Excellence Strategy EXC-2111 (Project No.~390814868), and are part of the Munich Quantum Valley, supported by the Bavarian state government through the Hightech Agenda Bayern Plus. H.-H.T. is supported by the Deutsche Forschungsgemeinschaft through project A06 of SFB 1143 (project-id 247310070).

\appendix
\section{Particle-number-conserving fermionic Gaussian states}
\label{app:U1-Gaussian-formalism}

In this Appendix, we provide further details on particle-number-conserving fermionic Gaussian states.
The proof of Eq.~\eqref{eq:GfPEPS-correlation-matrix-2} is also given.

To begin with, we briefly review the formalism of fermionic Gaussian states~\cite{Bravyi2005}.
Consider a system of $n$ fermionic modes with creation (annihilation) operators $c_{j}^{\dag }$ ($c_{j}$), $j=1,\ldots,n$.
Their linear combinations
\begin{equation}
\gamma _{2j-1}=c_{j}^{\dag }+c_{j}, \quad \gamma_{2j}=(-i)(c_{j}^{\dag}-c_{j})
\label{eq:Majorana-Operators}
\end{equation}
define $2n$ Majorana operators satisfying $\{\gamma_{a},\gamma_{b}\}=2\delta_{ab}$ ($a,b=1,\ldots ,2n$).
The density operator $\rho$, for both pure and mixed states, can be written as a polynomial in $\gamma_{a}$:
\begin{equation}
\rho =\frac{1}{2^{n}}\left( \mathbbm{1}+\frac{i}{2}\gamma ^{T}G\gamma +\cdots \right) , 
\label{eq:Density-Operator}
\end{equation}
where $\gamma =(\gamma _{1},\gamma _{2},\ldots ,\gamma _{2n})^{T}$, $\mathbbm{1}$ is the identity operator in the $2^{n}$-dimensional Hilbert space, and the ellipsis stands for terms with more than two (but even number of) Majorana operators. 
The real skew-symmetric matrix $G$ encodes two-point correlators in $\rho $, i.e., $G_{ab}=\frac{i}{2}\mathrm{tr}(\rho [\gamma_{a},\gamma_{b}])$.
This so-called correlation matrix $G$ satisfies $G^{T}G\leq \mathbbm{1}_{2n}$ with $\mathbbm{1}_{2n}$ being the $2n \times 2n$ identity matrix, and $G^{T}G=\mathbbm{1}_{2n}$ is achieved if and only if $\rho $ describes a pure state.
An operational definition of \emph{fermionic Gaussian state} is through the Grassmann representation of $\rho$ in Eq.~\eqref{eq:Density-Operator}: If one replaces each $\gamma_{a}$ by its corresponding Grassmann variable $\theta_{a}$ (and the identity operator $I$ by $1$), the Grassmann representation for a fermionic Gaussian state $\rho$, denoted by $\omega (\rho ,\theta )$, takes the following Gaussian form:
\begin{equation}
\omega (\rho ,\theta )=\frac{1}{2^{n}}\exp \left( \frac{i}{2}\theta
^{T}G\theta \right) ,
\label{eq:Grassmann-Density-Operator-1}
\end{equation}
where $\theta=(\theta_{1}, \theta_{2}, \ldots, \theta _{2n})^T$. The expansion of the exponential in Eq.~\eqref{eq:Grassmann-Density-Operator-1} gives all multipoint correlators in $\rho$, which are just coefficients of the respective Grassmann monomials and can be easily verified to be determined by Wick's theorem.

For our purpose, we would like to restrict ourselves to fermionic Gaussian states with a \emph{fixed particle number}. That means, the density operator $\rho$ in Eq.~\eqref{eq:Density-Operator}, apart from being Gaussian, should also commute with the total fermion number operator
\begin{align}
N =\sum_{j=1}^{n}c_{j}^{\dag }c_{j} = \frac{n}{2}\mathbbm{1}-\frac{i}{2}\gamma ^{T}Q\gamma
\label{eq:Number-Operator}
\end{align}
with $Q=\mathbbm{1}_{n}\otimes i\sigma^{y}$.
For $[\rho ,N]=0$ to hold, the correlation matrix $G$ must take the following form:
\begin{equation}
G=G_{1}\otimes \mathbbm{1}_{2}+G_{2}\otimes i\sigma ^{y},
\label{eq:Real-G}
\end{equation}
where the $n\times n$ matrix $G_{1}$ ($G_{2}$) is real and skew-symmetric (symmetric).
This structure can also be seen by requiring that there are no pairing correlations in $\rho$, i.e., $\mathrm{tr}(\rho c_{i}^{\dag}c_{j}^{\dag })=\mathrm{tr}(\rho c_{i}c_{j})=0$ $\forall i,j$.
It is then more natural to use a $n\times n$ ``complex'' correlation matrix
\begin{equation}
\mathcal{C}_{ij}\equiv 2\mathrm{tr}(\rho c_{i}^{\dag }c_{j})-\delta _{ij},
\label{eq:Complex-C}
\end{equation}
which relates to the ``real'' one in Eq.~\eqref{eq:Real-G} via $\mathcal{C}=-G_{2}-iG_{1}$.
The complex correlation matrix $\mathcal{C}$ is Hermitian and has eigenvalues $\lambda_{q}\in
[-1,1]$ $\forall q=1,\ldots ,n$. 
If all $\lambda _{q}=\pm 1$, $\rho$ is a pure state and the complex correlation matrix satisfies $\mathcal{C}^{-1}=\mathcal{C}$.
The diagonalization of $\mathcal{C}$ with a unitary matrix $U$ via $(U^{\dagger}\mathcal{C}U)_{qq'}=\lambda_{q}\delta _{qq'}$ defines the eigenmodes of $\rho$:
\begin{equation}
d^{\dag}_{q}=\sum_{j=1}^{n}U^{\dagger}_{qj}c^{\dag}_{j} .
\end{equation}
This brings $\rho$ into a simple form
\begin{equation}
\rho = \prod_{q=1}^{n}\left(\frac{1+\lambda_{q}}{2}d_{q}^{\dag}d_{q} +\frac{1-\lambda_{q}}{2}d_{q}d_{q}^{\dag}\right) ,
\end{equation}
where $d_{q}^{\dag}d_{q}$ ($d_{q}d_{q}^{\dag }$) is a projector onto an occupied (empty) state of $d_{q}$-mode.
Thus, the eigenmodes $d_{q}^{\dag}$ associated with $\lambda_{q}=1$ ($-1$) correspond to occupied (empty) single-particle orbitals in $\rho$.
For a pure state $\rho$, the number of eigenvalues with $\lambda_{q}=1$ is equal to the total number of occupied fermions.

The Grassmann representation is a convenient tool for fermionic Gaussian states~\cite{Bravyi2005}.
To adjust this tool for the particle-number-conserving case, we define $n$ pairs of ``complex'' Grassmann variables
\begin{equation}
\bar{\xi}_{j}=\frac{1}{\sqrt{2}}(\theta _{2j-1}-i\theta _{2j}),\quad \xi_{j}=\frac{1}{\sqrt{2}}(\theta_{2j-1}+i\theta _{2j})
\label{eq:Complex-Grassmann}
\end{equation}
with $j=1,\ldots,n$.
After substituting them into Eq.~\eqref{eq:Grassmann-Density-Operator-1} and using the relation between real and complex correlation matrices [Eqs.~\eqref{eq:Real-G} and \eqref{eq:Complex-C}], we arrive at the following ``complex'' Grassmann representation of $\rho$:
\begin{equation}
\omega (\rho ,\bar{\xi},\xi )=\frac{1}{2^{n}}\exp (-\bar{\xi}^{T}\mathcal{C}\xi),
\label{eq:Grassmann-Density-Operator-2}
\end{equation}
where $\xi =(\xi _{1},\xi _{2},\ldots ,\xi _{n})^{T}$ and $\bar{\xi}$ is similarly defined.

For constructing GfPEPS, one needs to deal with both \textit{physical} and \textit{virtual} fermionic modes.
Let us consider $n$ physical and $m$ virtual modes whose creation operators are $c_{j}^{\dag }$ $(j=1,\ldots ,n)$ and $b_{l}^{\dag }$ $(l=1,\ldots ,m)$, respectively.
The input is a Gaussian density operator $\rho _{\mathrm{in}}$ residing solely in the virtual Hilbert space.
The GfPEPS projector, formulated as another Gaussian density operator $\rho_{\mathrm{T}}$, lives in the composite Hilbert space of physical and virtual modes.
The Gaussian density operator of GfPEPS is written as
\begin{equation}
\rho_{\mathrm{out}}\propto \mathrm{tr}_{\mathrm{v}}(\rho_{\mathrm{T}}\rho_{\mathrm{in}}),
\end{equation}
where the partial trace $\mathrm{tr}_{\mathrm{v}}$ is with respect to the virtual Hilbert space.
It is shown in Ref.~\cite{Yang2022} that the correlation matrix of $\rho_{\mathrm{out}}$ can be calculated by using the Grassmann representation of $\mathrm{tr}_{\mathrm{v}}(\rho_{\mathrm{T}}\rho _{\mathrm{in}})$.
We can readily generalize this approach to the particle-number-conserving setting by converting ``real'' Grassmann variables to ``complex'' ones [see Eq.~\eqref{eq:Complex-Grassmann}] and obtain
\begin{align}
\mathrm{tr}_{\mathrm{v}}(\rho_{\mathrm{T}}\rho _{\mathrm{in}})(\bar{\xi},\xi )
&=2^{m}\int D\bar{\eta}D\eta D\bar{\mu}D\mu \; e^{\bar{\eta}^{T}\mu -\bar{\mu}^{T}\eta} \nonumber \\
& \phantom{=}\, \times \omega (\rho_{\mathrm{T}},\bar{\xi},\xi ,\bar{\eta},\eta )\omega (\rho _{\mathrm{in}},\bar{\mu},\mu ),
\label{eq:Partial-Trace}
\end{align}
where $\bar{\xi},\xi $ ($\bar{\eta},\eta ,\bar{\mu},\mu $) are Grassmann variables for physical (virtual) modes and $D\bar{\eta}D\eta =\mathrm{d}\bar{\eta}_{1}\mathrm{d}\eta _{1}\cdots \mathrm{d}\bar{\eta}_{m}\mathrm{d}\eta_{m}$ (similar for $D\bar{\mu}D\mu $).
By using the Grassmann representation of $\rho_{\mathrm{T}}$ and $\rho_{\mathrm{in}}$, namely,
\begin{equation*}
\begin{split}
\omega (\rho_{\mathrm{T}},\bar{\xi},\xi ,\bar{\eta},\eta ) 
&=\frac{1}{2^{n+m}}
\exp \left[ -(\bar{\xi}^{T} \; \bar{\eta}^{T})
\begin{pmatrix}
A & B \\
B^{\dag } & D
\end{pmatrix}
\begin{pmatrix}
\xi  \\
\eta
\end{pmatrix}
\right] , \\
\omega (\rho _{\mathrm{in}},\bar{\mu},\mu )
&=\frac{1}{2^{m}}\exp (-\bar{\mu}^{T}\mathcal{C}_{\mathrm{in}}\mu ),
\end{split}
\end{equation*}
and performing Gaussian integrations in Eq.~\eqref{eq:Partial-Trace}, we obtain
\begin{align}
\mathrm{tr}_{\mathrm{v}}(\rho_{\mathrm{T}}\rho _{\mathrm{in}})(\bar{\xi},\xi)
&=\frac{1}{2^{n+m}}\mathrm{\det }(\mathcal{C}_{\mathrm{in}})\det (D+\mathcal{C}_{\mathrm{in}}^{-1}) \nonumber \\
&\phantom{=} \, \times \exp( -\bar{\xi}^{T}\mathcal{C}_{\mathrm{out}}\xi ),
\end{align}
where the correlation matrix of $\rho _{\mathrm{out}}$ reads
\begin{equation}
\mathcal{C}_{\mathrm{out}}=A-B(D+\mathcal{C}_{\mathrm{in}}^{-1})^{-1}B^{\dag},
\end{equation}
For U(1)-GfPEPS, $\rho _{\mathrm{in}}$ is a pure state and satisfies $\mathcal{C}_{\mathrm{in}}^{-1}=\mathcal{C}_{\mathrm{in}}$. This completes the proof of Eq.~\eqref{eq:GfPEPS-correlation-matrix-2}.

\bibliography{gpeps}

\begin{thebibliography}{71}%
\makeatletter
\providecommand \@ifxundefined [1]{%
 \@ifx{#1\undefined}
}%
\providecommand \@ifnum [1]{%
 \ifnum #1\expandafter \@firstoftwo
 \else \expandafter \@secondoftwo
 \fi
}%
\providecommand \@ifx [1]{%
 \ifx #1\expandafter \@firstoftwo
 \else \expandafter \@secondoftwo
 \fi
}%
\providecommand \natexlab [1]{#1}%
\providecommand \enquote  [1]{``#1''}%
\providecommand \bibnamefont  [1]{#1}%
\providecommand \bibfnamefont [1]{#1}%
\providecommand \citenamefont [1]{#1}%
\providecommand \href@noop [0]{\@secondoftwo}%
\providecommand \href [0]{\begingroup \@sanitize@url \@href}%
\providecommand \@href[1]{\@@startlink{#1}\@@href}%
\providecommand \@@href[1]{\endgroup#1\@@endlink}%
\providecommand \@sanitize@url [0]{\catcode `\\12\catcode `\$12\catcode
  `\&12\catcode `\#12\catcode `\^12\catcode `\_12\catcode `\%12\relax}%
\providecommand \@@startlink[1]{}%
\providecommand \@@endlink[0]{}%
\providecommand \url  [0]{\begingroup\@sanitize@url \@url }%
\providecommand \@url [1]{\endgroup\@href {#1}{\urlprefix }}%
\providecommand \urlprefix  [0]{URL }%
\providecommand \Eprint [0]{\href }%
\providecommand \doibase [0]{http://dx.doi.org/}%
\providecommand \selectlanguage [0]{\@gobble}%
\providecommand \bibinfo  [0]{\@secondoftwo}%
\providecommand \bibfield  [0]{\@secondoftwo}%
\providecommand \translation [1]{[#1]}%
\providecommand \BibitemOpen [0]{}%
\providecommand \bibitemStop [0]{}%
\providecommand \bibitemNoStop [0]{.\EOS\space}%
\providecommand \EOS [0]{\spacefactor3000\relax}%
\providecommand \BibitemShut  [1]{\csname bibitem#1\endcsname}%
\let\auto@bib@innerbib\@empty
\bibitem [{\citenamefont {Gutzwiller}(1963)}]{Gutzwiller1963}%
  \BibitemOpen
  \bibfield  {author} {\bibinfo {author} {\bibfnamefont {M.~C.}\ \bibnamefont
  {Gutzwiller}},\ }\href {\doibase 10.1103/PhysRevLett.10.159} {\bibfield
  {journal} {\bibinfo  {journal} {Phys. Rev. Lett.}\ }\textbf {\bibinfo
  {volume} {10}},\ \bibinfo {pages} {159} (\bibinfo {year} {1963})}\BibitemShut
  {NoStop}%
\bibitem [{\citenamefont {Gutzwiller}(1965)}]{Gutzwiller1965}%
  \BibitemOpen
  \bibfield  {author} {\bibinfo {author} {\bibfnamefont {M.~C.}\ \bibnamefont
  {Gutzwiller}},\ }\href {\doibase 10.1103/PhysRev.137.A1726} {\bibfield
  {journal} {\bibinfo  {journal} {Phys. Rev.}\ }\textbf {\bibinfo {volume}
  {137}},\ \bibinfo {pages} {A1726} (\bibinfo {year} {1965})}\BibitemShut
  {NoStop}%
\bibitem [{\citenamefont {Anderson}(1987)}]{Anderson1987}%
  \BibitemOpen
  \bibfield  {author} {\bibinfo {author} {\bibfnamefont {P.~W.}\ \bibnamefont
  {Anderson}},\ }\href {\doibase 10.1126/science.235.4793.1196} {\bibfield
  {journal} {\bibinfo  {journal} {Science}\ }\textbf {\bibinfo {volume}
  {235}},\ \bibinfo {pages} {1196} (\bibinfo {year} {1987})}\BibitemShut
  {NoStop}%
\bibitem [{\citenamefont {Haldane}(1988)}]{Haldane1988}%
  \BibitemOpen
  \bibfield  {author} {\bibinfo {author} {\bibfnamefont {F.~D.~M.}\
  \bibnamefont {Haldane}},\ }\href
  {https://journals.aps.org/prl/abstract/10.1103/PhysRevLett.60.635} {\bibfield
   {journal} {\bibinfo  {journal} {Phys. Rev. Lett.}\ }\textbf {\bibinfo
  {volume} {60}},\ \bibinfo {pages} {635} (\bibinfo {year} {1988})}\BibitemShut
  {NoStop}%
\bibitem [{\citenamefont {Shastry}(1988)}]{Shastry1988}%
  \BibitemOpen
  \bibfield  {author} {\bibinfo {author} {\bibfnamefont {B.~S.}\ \bibnamefont
  {Shastry}},\ }\href
  {https://journals.aps.org/prl/abstract/10.1103/PhysRevLett.60.639} {\bibfield
   {journal} {\bibinfo  {journal} {Phys. Rev. Lett.}\ }\textbf {\bibinfo
  {volume} {60}},\ \bibinfo {pages} {639} (\bibinfo {year} {1988})}\BibitemShut
  {NoStop}%
\bibitem [{\citenamefont {Kitaev}(2006)}]{Kitaev2006}%
  \BibitemOpen
  \bibfield  {author} {\bibinfo {author} {\bibfnamefont {A.}~\bibnamefont
  {Kitaev}},\ }\href {\doibase https://doi.org/10.1016/j.aop.2005.10.005}
  {\bibfield  {journal} {\bibinfo  {journal} {Ann. Phys.}\ }\textbf {\bibinfo
  {volume} {321}},\ \bibinfo {pages} {2} (\bibinfo {year} {2006})}\BibitemShut
  {NoStop}%
\bibitem [{\citenamefont {Gros}\ \emph {et~al.}(1987)\citenamefont {Gros},
  \citenamefont {Joynt},\ and\ \citenamefont {Rice}}]{Gros1987}%
  \BibitemOpen
  \bibfield  {author} {\bibinfo {author} {\bibfnamefont {C.}~\bibnamefont
  {Gros}}, \bibinfo {author} {\bibfnamefont {R.}~\bibnamefont {Joynt}}, \ and\
  \bibinfo {author} {\bibfnamefont {T.~M.}\ \bibnamefont {Rice}},\ }\href
  {\doibase 10.1103/PhysRevB.36.381} {\bibfield  {journal} {\bibinfo  {journal}
  {Phys. Rev. B}\ }\textbf {\bibinfo {volume} {36}},\ \bibinfo {pages} {381}
  (\bibinfo {year} {1987})}\BibitemShut {NoStop}%
\bibitem [{\citenamefont {Yokoyama}\ and\ \citenamefont
  {Shiba}(1987)}]{Yokoyama1987}%
  \BibitemOpen
  \bibfield  {author} {\bibinfo {author} {\bibfnamefont {H.}~\bibnamefont
  {Yokoyama}}\ and\ \bibinfo {author} {\bibfnamefont {H.}~\bibnamefont
  {Shiba}},\ }\href {https://doi.org/10.1143/JPSJ.56.1490} {\bibfield
  {journal} {\bibinfo  {journal} {J. Phys. Soc. Jpn}\ }\textbf {\bibinfo
  {volume} {56}},\ \bibinfo {pages} {1490} (\bibinfo {year}
  {1987})}\BibitemShut {NoStop}%
\bibitem [{\citenamefont {Gros}(1989)}]{Gros1989}%
  \BibitemOpen
  \bibfield  {author} {\bibinfo {author} {\bibfnamefont {C.}~\bibnamefont
  {Gros}},\ }\href {\doibase https://doi.org/10.1016/0003-4916(89)90077-8}
  {\bibfield  {journal} {\bibinfo  {journal} {Ann. Phys.}\ }\textbf {\bibinfo
  {volume} {189}},\ \bibinfo {pages} {53} (\bibinfo {year} {1989})}\BibitemShut
  {NoStop}%
\bibitem [{\citenamefont {Fishman}\ and\ \citenamefont
  {White}(2015)}]{Fishman2015}%
  \BibitemOpen
  \bibfield  {author} {\bibinfo {author} {\bibfnamefont {M.~T.}\ \bibnamefont
  {Fishman}}\ and\ \bibinfo {author} {\bibfnamefont {S.~R.}\ \bibnamefont
  {White}},\ }\href {\doibase 10.1103/PhysRevB.92.075132} {\bibfield  {journal}
  {\bibinfo  {journal} {Phys. Rev. B}\ }\textbf {\bibinfo {volume} {92}},\
  \bibinfo {pages} {075132} (\bibinfo {year} {2015})}\BibitemShut {NoStop}%
\bibitem [{\citenamefont {Wu}\ \emph {et~al.}(2020)\citenamefont {Wu},
  \citenamefont {Wang},\ and\ \citenamefont {Tu}}]{Wu2020}%
  \BibitemOpen
  \bibfield  {author} {\bibinfo {author} {\bibfnamefont {Y.-H.}\ \bibnamefont
  {Wu}}, \bibinfo {author} {\bibfnamefont {L.}~\bibnamefont {Wang}}, \ and\
  \bibinfo {author} {\bibfnamefont {H.-H.}\ \bibnamefont {Tu}},\ }\href
  {\doibase 10.1103/PhysRevLett.124.246401} {\bibfield  {journal} {\bibinfo
  {journal} {Phys. Rev. Lett.}\ }\textbf {\bibinfo {volume} {124}},\ \bibinfo
  {pages} {246401} (\bibinfo {year} {2020})}\BibitemShut {NoStop}%
\bibitem [{\citenamefont {Jin}\ \emph {et~al.}(2020)\citenamefont {Jin},
  \citenamefont {Tu},\ and\ \citenamefont {Zhou}}]{Jin2020}%
  \BibitemOpen
  \bibfield  {author} {\bibinfo {author} {\bibfnamefont {H.-K.}\ \bibnamefont
  {Jin}}, \bibinfo {author} {\bibfnamefont {H.-H.}\ \bibnamefont {Tu}}, \ and\
  \bibinfo {author} {\bibfnamefont {Y.}~\bibnamefont {Zhou}},\ }\href
  {https://link.aps.org/doi/10.1103/PhysRevB.101.165135} {\bibfield  {journal}
  {\bibinfo  {journal} {Phys. Rev. B}\ }\textbf {\bibinfo {volume} {101}},\
  \bibinfo {pages} {165135} (\bibinfo {year} {2020})}\BibitemShut {NoStop}%
\bibitem [{\citenamefont {Aghaei}\ \emph {et~al.}()\citenamefont {Aghaei},
  \citenamefont {Bauer}, \citenamefont {Shtengel},\ and\ \citenamefont
  {Mishmash}}]{Aghaei2020}%
  \BibitemOpen
  \bibfield  {author} {\bibinfo {author} {\bibfnamefont {A.~M.}\ \bibnamefont
  {Aghaei}}, \bibinfo {author} {\bibfnamefont {B.}~\bibnamefont {Bauer}},
  \bibinfo {author} {\bibfnamefont {K.}~\bibnamefont {Shtengel}}, \ and\
  \bibinfo {author} {\bibfnamefont {R.~V.}\ \bibnamefont {Mishmash}},\
  }\href@noop {} {\ }\Eprint {http://arxiv.org/abs/arXiv:2009.12435}
  {arXiv:2009.12435} \BibitemShut {NoStop}%
\bibitem [{\citenamefont {Petrica}\ \emph {et~al.}(2021)\citenamefont
  {Petrica}, \citenamefont {Zheng}, \citenamefont {Chan},\ and\ \citenamefont
  {Clark}}]{Petrica2021}%
  \BibitemOpen
  \bibfield  {author} {\bibinfo {author} {\bibfnamefont {G.}~\bibnamefont
  {Petrica}}, \bibinfo {author} {\bibfnamefont {B.-X.}\ \bibnamefont {Zheng}},
  \bibinfo {author} {\bibfnamefont {G.~K.-L.}\ \bibnamefont {Chan}}, \ and\
  \bibinfo {author} {\bibfnamefont {B.~K.}\ \bibnamefont {Clark}},\ }\href
  {\doibase 10.1103/PhysRevB.103.125161} {\bibfield  {journal} {\bibinfo
  {journal} {Phys. Rev. B}\ }\textbf {\bibinfo {volume} {103}},\ \bibinfo
  {pages} {125161} (\bibinfo {year} {2021})}\BibitemShut {NoStop}%
\bibitem [{\citenamefont {Jones}\ \emph {et~al.}(2021)\citenamefont {Jones},
  \citenamefont {Bibo}, \citenamefont {Jobst}, \citenamefont {Pollmann},
  \citenamefont {Smith},\ and\ \citenamefont {Verresen}}]{Jones2021a}%
  \BibitemOpen
  \bibfield  {author} {\bibinfo {author} {\bibfnamefont {N.~G.}\ \bibnamefont
  {Jones}}, \bibinfo {author} {\bibfnamefont {J.}~\bibnamefont {Bibo}},
  \bibinfo {author} {\bibfnamefont {B.}~\bibnamefont {Jobst}}, \bibinfo
  {author} {\bibfnamefont {F.}~\bibnamefont {Pollmann}}, \bibinfo {author}
  {\bibfnamefont {A.}~\bibnamefont {Smith}}, \ and\ \bibinfo {author}
  {\bibfnamefont {R.}~\bibnamefont {Verresen}},\ }\href {\doibase
  10.1103/PhysRevResearch.3.033265} {\bibfield  {journal} {\bibinfo  {journal}
  {Phys. Rev. Research}\ }\textbf {\bibinfo {volume} {3}},\ \bibinfo {pages}
  {033265} (\bibinfo {year} {2021})}\BibitemShut {NoStop}%
\bibitem [{\citenamefont {Jin}\ \emph {et~al.}(2022{\natexlab{a}})\citenamefont
  {Jin}, \citenamefont {Sun}, \citenamefont {Zhou},\ and\ \citenamefont
  {Tu}}]{Jin2022a}%
  \BibitemOpen
  \bibfield  {author} {\bibinfo {author} {\bibfnamefont {H.-K.}\ \bibnamefont
  {Jin}}, \bibinfo {author} {\bibfnamefont {R.-Y.}\ \bibnamefont {Sun}},
  \bibinfo {author} {\bibfnamefont {Y.}~\bibnamefont {Zhou}}, \ and\ \bibinfo
  {author} {\bibfnamefont {H.-H.}\ \bibnamefont {Tu}},\ }\href {\doibase
  10.1103/PhysRevB.105.L081101} {\bibfield  {journal} {\bibinfo  {journal}
  {Phys. Rev. B}\ }\textbf {\bibinfo {volume} {105}},\ \bibinfo {pages}
  {L081101} (\bibinfo {year} {2022}{\natexlab{a}})}\BibitemShut {NoStop}%
\bibitem [{\citenamefont {White}(1992)}]{White1992}%
  \BibitemOpen
  \bibfield  {author} {\bibinfo {author} {\bibfnamefont {S.~R.}\ \bibnamefont
  {White}},\ }\href {\doibase 10.1103/PhysRevLett.69.2863} {\bibfield
  {journal} {\bibinfo  {journal} {Phys. Rev. Lett.}\ }\textbf {\bibinfo
  {volume} {69}},\ \bibinfo {pages} {2863} (\bibinfo {year}
  {1992})}\BibitemShut {NoStop}%
\bibitem [{\citenamefont {\"Ostlund}\ and\ \citenamefont
  {Rommer}(1995)}]{Ostlund1995}%
  \BibitemOpen
  \bibfield  {author} {\bibinfo {author} {\bibfnamefont {S.}~\bibnamefont
  {\"Ostlund}}\ and\ \bibinfo {author} {\bibfnamefont {S.}~\bibnamefont
  {Rommer}},\ }\href {\doibase 10.1103/PhysRevLett.75.3537} {\bibfield
  {journal} {\bibinfo  {journal} {Phys. Rev. Lett.}\ }\textbf {\bibinfo
  {volume} {75}},\ \bibinfo {pages} {3537} (\bibinfo {year}
  {1995})}\BibitemShut {NoStop}%
\bibitem [{\citenamefont {Verstraete}\ \emph {et~al.}(2008)\citenamefont
  {Verstraete}, \citenamefont {Murg},\ and\ \citenamefont
  {Cirac}}]{Verstraete2008}%
  \BibitemOpen
  \bibfield  {author} {\bibinfo {author} {\bibfnamefont {F.}~\bibnamefont
  {Verstraete}}, \bibinfo {author} {\bibfnamefont {V.}~\bibnamefont {Murg}}, \
  and\ \bibinfo {author} {\bibfnamefont {J.~I.}\ \bibnamefont {Cirac}},\ }\href
  {\doibase 10.1080/14789940801912366} {\bibfield  {journal} {\bibinfo
  {journal} {Adv. Phys.}\ }\textbf {\bibinfo {volume} {57}},\ \bibinfo {pages}
  {143} (\bibinfo {year} {2008})}\BibitemShut {NoStop}%
\bibitem [{\citenamefont {Schollw\"ock}(2011)}]{Schollwoeck2011}%
  \BibitemOpen
  \bibfield  {author} {\bibinfo {author} {\bibfnamefont {U.}~\bibnamefont
  {Schollw\"ock}},\ }\href {\doibase https://doi.org/10.1016/j.aop.2010.09.012}
  {\bibfield  {journal} {\bibinfo  {journal} {Ann. Phys.}\ }\textbf {\bibinfo
  {volume} {326}},\ \bibinfo {pages} {96 } (\bibinfo {year}
  {2011})}\BibitemShut {NoStop}%
\bibitem [{\citenamefont {Jin}\ \emph {et~al.}(2021)\citenamefont {Jin},
  \citenamefont {Tu},\ and\ \citenamefont {Zhou}}]{Jin2021}%
  \BibitemOpen
  \bibfield  {author} {\bibinfo {author} {\bibfnamefont {H.-K.}\ \bibnamefont
  {Jin}}, \bibinfo {author} {\bibfnamefont {H.-H.}\ \bibnamefont {Tu}}, \ and\
  \bibinfo {author} {\bibfnamefont {Y.}~\bibnamefont {Zhou}},\ }\href {\doibase
  10.1103/PhysRevB.104.L020409} {\bibfield  {journal} {\bibinfo  {journal}
  {Phys. Rev. B}\ }\textbf {\bibinfo {volume} {104}},\ \bibinfo {pages}
  {L020409} (\bibinfo {year} {2021})}\BibitemShut {NoStop}%
\bibitem [{\citenamefont {Chen}\ \emph {et~al.}(2021)\citenamefont {Chen},
  \citenamefont {Li}, \citenamefont {Nataf}, \citenamefont {Capponi},
  \citenamefont {Mambrini}, \citenamefont {Totsuka}, \citenamefont {Tu},
  \citenamefont {Weichselbaum}, \citenamefont {von Delft},\ and\ \citenamefont
  {Poilblanc}}]{Chen2021}%
  \BibitemOpen
  \bibfield  {author} {\bibinfo {author} {\bibfnamefont {J.-Y.}\ \bibnamefont
  {Chen}}, \bibinfo {author} {\bibfnamefont {J.-W.}\ \bibnamefont {Li}},
  \bibinfo {author} {\bibfnamefont {P.}~\bibnamefont {Nataf}}, \bibinfo
  {author} {\bibfnamefont {S.}~\bibnamefont {Capponi}}, \bibinfo {author}
  {\bibfnamefont {M.}~\bibnamefont {Mambrini}}, \bibinfo {author}
  {\bibfnamefont {K.}~\bibnamefont {Totsuka}}, \bibinfo {author} {\bibfnamefont
  {H.-H.}\ \bibnamefont {Tu}}, \bibinfo {author} {\bibfnamefont
  {A.}~\bibnamefont {Weichselbaum}}, \bibinfo {author} {\bibfnamefont
  {J.}~\bibnamefont {von Delft}}, \ and\ \bibinfo {author} {\bibfnamefont
  {D.}~\bibnamefont {Poilblanc}},\ }\href {\doibase
  10.1103/PhysRevB.104.235104} {\bibfield  {journal} {\bibinfo  {journal}
  {Phys. Rev. B}\ }\textbf {\bibinfo {volume} {104}},\ \bibinfo {pages}
  {235104} (\bibinfo {year} {2021})}\BibitemShut {NoStop}%
\bibitem [{\citenamefont {Jin}\ \emph {et~al.}(2022{\natexlab{b}})\citenamefont
  {Jin}, \citenamefont {Sun}, \citenamefont {Tu},\ and\ \citenamefont
  {Zhou}}]{Jin2022b}%
  \BibitemOpen
  \bibfield  {author} {\bibinfo {author} {\bibfnamefont {H.-K.}\ \bibnamefont
  {Jin}}, \bibinfo {author} {\bibfnamefont {R.-Y.}\ \bibnamefont {Sun}},
  \bibinfo {author} {\bibfnamefont {H.-H.}\ \bibnamefont {Tu}}, \ and\ \bibinfo
  {author} {\bibfnamefont {Y.}~\bibnamefont {Zhou}},\ }\href
  {https://doi.org/10.1016/j.scib.2022.03.004} {\bibfield  {journal} {\bibinfo
  {journal} {Sci. Bull.}\ } (\bibinfo {year} {2022}{\natexlab{b}})}\BibitemShut
  {NoStop}%
\bibitem [{\citenamefont {Sun}\ \emph {et~al.}()\citenamefont {Sun},
  \citenamefont {Jin}, \citenamefont {Tu},\ and\ \citenamefont
  {Zhou}}]{Sun2022}%
  \BibitemOpen
  \bibfield  {author} {\bibinfo {author} {\bibfnamefont {R.-Y.}\ \bibnamefont
  {Sun}}, \bibinfo {author} {\bibfnamefont {H.-K.}\ \bibnamefont {Jin}},
  \bibinfo {author} {\bibfnamefont {H.-H.}\ \bibnamefont {Tu}}, \ and\ \bibinfo
  {author} {\bibfnamefont {Y.}~\bibnamefont {Zhou}},\ }\href@noop {} {\
  }\Eprint {http://arxiv.org/abs/arXiv:2203.07321} {arXiv:2203.07321}
  \BibitemShut {NoStop}%
\bibitem [{\citenamefont {Verstraete}\ and\ \citenamefont
  {Cirac}()}]{Verstraete2004}%
  \BibitemOpen
  \bibfield  {author} {\bibinfo {author} {\bibfnamefont {F.}~\bibnamefont
  {Verstraete}}\ and\ \bibinfo {author} {\bibfnamefont {J.~I.}\ \bibnamefont
  {Cirac}},\ }\href {https://arxiv.org/abs/cond-mat/0407066} {\ }\Eprint
  {http://arxiv.org/abs/arXiv: cond-mat/0407066} {arXiv: cond-mat/0407066}
  \BibitemShut {NoStop}%
\bibitem [{\citenamefont {Jiang}\ \emph {et~al.}(2008)\citenamefont {Jiang},
  \citenamefont {Weng},\ and\ \citenamefont {Xiang}}]{Jiang2008}%
  \BibitemOpen
  \bibfield  {author} {\bibinfo {author} {\bibfnamefont {H.~C.}\ \bibnamefont
  {Jiang}}, \bibinfo {author} {\bibfnamefont {Z.~Y.}\ \bibnamefont {Weng}}, \
  and\ \bibinfo {author} {\bibfnamefont {T.}~\bibnamefont {Xiang}},\ }\href
  {\doibase 10.1103/PhysRevLett.101.090603} {\bibfield  {journal} {\bibinfo
  {journal} {Phys. Rev. Lett.}\ }\textbf {\bibinfo {volume} {101}},\ \bibinfo
  {pages} {090603} (\bibinfo {year} {2008})}\BibitemShut {NoStop}%
\bibitem [{\citenamefont {Jordan}\ \emph {et~al.}(2008)\citenamefont {Jordan},
  \citenamefont {Or\'us}, \citenamefont {Vidal}, \citenamefont {Verstraete},\
  and\ \citenamefont {Cirac}}]{Jordan2008}%
  \BibitemOpen
  \bibfield  {author} {\bibinfo {author} {\bibfnamefont {J.}~\bibnamefont
  {Jordan}}, \bibinfo {author} {\bibfnamefont {R.}~\bibnamefont {Or\'us}},
  \bibinfo {author} {\bibfnamefont {G.}~\bibnamefont {Vidal}}, \bibinfo
  {author} {\bibfnamefont {F.}~\bibnamefont {Verstraete}}, \ and\ \bibinfo
  {author} {\bibfnamefont {J.~I.}\ \bibnamefont {Cirac}},\ }\href {\doibase
  10.1103/PhysRevLett.101.250602} {\bibfield  {journal} {\bibinfo  {journal}
  {Phys. Rev. Lett.}\ }\textbf {\bibinfo {volume} {101}},\ \bibinfo {pages}
  {250602} (\bibinfo {year} {2008})}\BibitemShut {NoStop}%
\bibitem [{\citenamefont {Corboz}(2016)}]{Corboz2016}%
  \BibitemOpen
  \bibfield  {author} {\bibinfo {author} {\bibfnamefont {P.}~\bibnamefont
  {Corboz}},\ }\href {\doibase 10.1103/PhysRevB.94.035133} {\bibfield
  {journal} {\bibinfo  {journal} {Phys. Rev. B}\ }\textbf {\bibinfo {volume}
  {94}},\ \bibinfo {pages} {035133} (\bibinfo {year} {2016})}\BibitemShut
  {NoStop}%
\bibitem [{\citenamefont {Vanderstraeten}\ \emph {et~al.}(2016)\citenamefont
  {Vanderstraeten}, \citenamefont {Haegeman}, \citenamefont {Corboz},\ and\
  \citenamefont {Verstraete}}]{Vanderstraeten2016}%
  \BibitemOpen
  \bibfield  {author} {\bibinfo {author} {\bibfnamefont {L.}~\bibnamefont
  {Vanderstraeten}}, \bibinfo {author} {\bibfnamefont {J.}~\bibnamefont
  {Haegeman}}, \bibinfo {author} {\bibfnamefont {P.}~\bibnamefont {Corboz}}, \
  and\ \bibinfo {author} {\bibfnamefont {F.}~\bibnamefont {Verstraete}},\
  }\href {\doibase 10.1103/PhysRevB.94.155123} {\bibfield  {journal} {\bibinfo
  {journal} {Phys. Rev. B}\ }\textbf {\bibinfo {volume} {94}},\ \bibinfo
  {pages} {155123} (\bibinfo {year} {2016})}\BibitemShut {NoStop}%
\bibitem [{\citenamefont {Liao}\ \emph {et~al.}(2019)\citenamefont {Liao},
  \citenamefont {Liu}, \citenamefont {Wang},\ and\ \citenamefont
  {Xiang}}]{Liao2019}%
  \BibitemOpen
  \bibfield  {author} {\bibinfo {author} {\bibfnamefont {H.-J.}\ \bibnamefont
  {Liao}}, \bibinfo {author} {\bibfnamefont {J.-G.}\ \bibnamefont {Liu}},
  \bibinfo {author} {\bibfnamefont {L.}~\bibnamefont {Wang}}, \ and\ \bibinfo
  {author} {\bibfnamefont {T.}~\bibnamefont {Xiang}},\ }\href {\doibase
  10.1103/PhysRevX.9.031041} {\bibfield  {journal} {\bibinfo  {journal} {Phys.
  Rev. X}\ }\textbf {\bibinfo {volume} {9}},\ \bibinfo {pages} {031041}
  (\bibinfo {year} {2019})}\BibitemShut {NoStop}%
\bibitem [{\citenamefont {Nishino}\ and\ \citenamefont
  {Okunishi}(1996)}]{Nishino1996}%
  \BibitemOpen
  \bibfield  {author} {\bibinfo {author} {\bibfnamefont {T.}~\bibnamefont
  {Nishino}}\ and\ \bibinfo {author} {\bibfnamefont {K.}~\bibnamefont
  {Okunishi}},\ }\href {\doibase 10.1143/JPSJ.65.891} {\bibfield  {journal}
  {\bibinfo  {journal} {J. Phys. Soc. Jpn}\ }\textbf {\bibinfo {volume} {65}},\
  \bibinfo {pages} {891} (\bibinfo {year} {1996})}\BibitemShut {NoStop}%
\bibitem [{\citenamefont {Levin}\ and\ \citenamefont {Nave}(2007)}]{Levin2007}%
  \BibitemOpen
  \bibfield  {author} {\bibinfo {author} {\bibfnamefont {M.}~\bibnamefont
  {Levin}}\ and\ \bibinfo {author} {\bibfnamefont {C.~P.}\ \bibnamefont
  {Nave}},\ }\href {\doibase 10.1103/PhysRevLett.99.120601} {\bibfield
  {journal} {\bibinfo  {journal} {Phys. Rev. Lett.}\ }\textbf {\bibinfo
  {volume} {99}},\ \bibinfo {pages} {120601} (\bibinfo {year}
  {2007})}\BibitemShut {NoStop}%
\bibitem [{\citenamefont {Or\'us}\ and\ \citenamefont
  {Vidal}(2009)}]{Orus2009}%
  \BibitemOpen
  \bibfield  {author} {\bibinfo {author} {\bibfnamefont {R.}~\bibnamefont
  {Or\'us}}\ and\ \bibinfo {author} {\bibfnamefont {G.}~\bibnamefont {Vidal}},\
  }\href {\doibase 10.1103/PhysRevB.80.094403} {\bibfield  {journal} {\bibinfo
  {journal} {Phys. Rev. B}\ }\textbf {\bibinfo {volume} {80}},\ \bibinfo
  {pages} {094403} (\bibinfo {year} {2009})}\BibitemShut {NoStop}%
\bibitem [{\citenamefont {Xie}\ \emph {et~al.}(2012)\citenamefont {Xie},
  \citenamefont {Chen}, \citenamefont {Qin}, \citenamefont {Zhu}, \citenamefont
  {Yang},\ and\ \citenamefont {Xiang}}]{Xie2012}%
  \BibitemOpen
  \bibfield  {author} {\bibinfo {author} {\bibfnamefont {Z.~Y.}\ \bibnamefont
  {Xie}}, \bibinfo {author} {\bibfnamefont {J.}~\bibnamefont {Chen}}, \bibinfo
  {author} {\bibfnamefont {M.~P.}\ \bibnamefont {Qin}}, \bibinfo {author}
  {\bibfnamefont {J.~W.}\ \bibnamefont {Zhu}}, \bibinfo {author} {\bibfnamefont
  {L.~P.}\ \bibnamefont {Yang}}, \ and\ \bibinfo {author} {\bibfnamefont
  {T.}~\bibnamefont {Xiang}},\ }\href {\doibase 10.1103/PhysRevB.86.045139}
  {\bibfield  {journal} {\bibinfo  {journal} {Phys. Rev. B}\ }\textbf {\bibinfo
  {volume} {86}},\ \bibinfo {pages} {045139} (\bibinfo {year}
  {2012})}\BibitemShut {NoStop}%
\bibitem [{\citenamefont {Fishman}\ \emph {et~al.}(2018)\citenamefont
  {Fishman}, \citenamefont {Vanderstraeten}, \citenamefont {Zauner-Stauber},
  \citenamefont {Haegeman},\ and\ \citenamefont {Verstraete}}]{Fishman2018}%
  \BibitemOpen
  \bibfield  {author} {\bibinfo {author} {\bibfnamefont {M.~T.}\ \bibnamefont
  {Fishman}}, \bibinfo {author} {\bibfnamefont {L.}~\bibnamefont
  {Vanderstraeten}}, \bibinfo {author} {\bibfnamefont {V.}~\bibnamefont
  {Zauner-Stauber}}, \bibinfo {author} {\bibfnamefont {J.}~\bibnamefont
  {Haegeman}}, \ and\ \bibinfo {author} {\bibfnamefont {F.}~\bibnamefont
  {Verstraete}},\ }\href {\doibase 10.1103/PhysRevB.98.235148} {\bibfield
  {journal} {\bibinfo  {journal} {Phys. Rev. B}\ }\textbf {\bibinfo {volume}
  {98}},\ \bibinfo {pages} {235148} (\bibinfo {year} {2018})}\BibitemShut
  {NoStop}%
\bibitem [{\citenamefont {Schuch}\ \emph {et~al.}(2010)\citenamefont {Schuch},
  \citenamefont {Cirac},\ and\ \citenamefont
  {P{\'e}rez-Garc{\'\i}a}}]{Schuch2010}%
  \BibitemOpen
  \bibfield  {author} {\bibinfo {author} {\bibfnamefont {N.}~\bibnamefont
  {Schuch}}, \bibinfo {author} {\bibfnamefont {I.}~\bibnamefont {Cirac}}, \
  and\ \bibinfo {author} {\bibfnamefont {D.}~\bibnamefont
  {P{\'e}rez-Garc{\'\i}a}},\ }\href {https://doi.org/10.1016/j.aop.2010.05.008}
  {\bibfield  {journal} {\bibinfo  {journal} {Ann. Phys.}\ }\textbf {\bibinfo
  {volume} {325}},\ \bibinfo {pages} {2153} (\bibinfo {year}
  {2010})}\BibitemShut {NoStop}%
\bibitem [{\citenamefont {Buerschaper}(2014)}]{Buerschaper2014}%
  \BibitemOpen
  \bibfield  {author} {\bibinfo {author} {\bibfnamefont {O.}~\bibnamefont
  {Buerschaper}},\ }\href {https://doi.org/10.1016/j.aop.2014.09.007}
  {\bibfield  {journal} {\bibinfo  {journal} {Ann. Phys.}\ }\textbf {\bibinfo
  {volume} {351}},\ \bibinfo {pages} {447} (\bibinfo {year}
  {2014})}\BibitemShut {NoStop}%
\bibitem [{\citenamefont {Williamson}\ \emph {et~al.}(2016)\citenamefont
  {Williamson}, \citenamefont {Bultinck}, \citenamefont {Mari\"en},
  \citenamefont {\ifmmode \mbox{\c{S}}\else
  \c{S}\fi{}ahino\ifmmode~\breve{g}\else \u{g}\fi{}lu}, \citenamefont
  {Haegeman},\ and\ \citenamefont {Verstraete}}]{Williamson2016}%
  \BibitemOpen
  \bibfield  {author} {\bibinfo {author} {\bibfnamefont {D.~J.}\ \bibnamefont
  {Williamson}}, \bibinfo {author} {\bibfnamefont {N.}~\bibnamefont
  {Bultinck}}, \bibinfo {author} {\bibfnamefont {M.}~\bibnamefont {Mari\"en}},
  \bibinfo {author} {\bibfnamefont {M.~B.}\ \bibnamefont {\ifmmode
  \mbox{\c{S}}\else \c{S}\fi{}ahino\ifmmode~\breve{g}\else \u{g}\fi{}lu}},
  \bibinfo {author} {\bibfnamefont {J.}~\bibnamefont {Haegeman}}, \ and\
  \bibinfo {author} {\bibfnamefont {F.}~\bibnamefont {Verstraete}},\ }\href
  {\doibase 10.1103/PhysRevB.94.205150} {\bibfield  {journal} {\bibinfo
  {journal} {Phys. Rev. B}\ }\textbf {\bibinfo {volume} {94}},\ \bibinfo
  {pages} {205150} (\bibinfo {year} {2016})}\BibitemShut {NoStop}%
\bibitem [{\citenamefont {{\c{S}}ahino{\u{g}}lu}\ \emph
  {et~al.}(2021)\citenamefont {{\c{S}}ahino{\u{g}}lu}, \citenamefont
  {Williamson}, \citenamefont {Bultinck}, \citenamefont {Mari{\"e}n},
  \citenamefont {Haegeman}, \citenamefont {Schuch},\ and\ \citenamefont
  {Verstraete}}]{Sahinouglu2021}%
  \BibitemOpen
  \bibfield  {author} {\bibinfo {author} {\bibfnamefont {M.~B.}\ \bibnamefont
  {{\c{S}}ahino{\u{g}}lu}}, \bibinfo {author} {\bibfnamefont {D.}~\bibnamefont
  {Williamson}}, \bibinfo {author} {\bibfnamefont {N.}~\bibnamefont
  {Bultinck}}, \bibinfo {author} {\bibfnamefont {M.}~\bibnamefont
  {Mari{\"e}n}}, \bibinfo {author} {\bibfnamefont {J.}~\bibnamefont
  {Haegeman}}, \bibinfo {author} {\bibfnamefont {N.}~\bibnamefont {Schuch}}, \
  and\ \bibinfo {author} {\bibfnamefont {F.}~\bibnamefont {Verstraete}},\
  }\href {https://doi.org/10.1007/s00023-020-00992-4} {\bibfield  {journal}
  {\bibinfo  {journal} {Ann. Henri Poincar{\'e}}\ }\textbf {\bibinfo {volume}
  {22}},\ \bibinfo {pages} {563} (\bibinfo {year} {2021})}\BibitemShut
  {NoStop}%
\bibitem [{\citenamefont {Kraus}\ \emph {et~al.}(2010)\citenamefont {Kraus},
  \citenamefont {Schuch}, \citenamefont {Verstraete},\ and\ \citenamefont
  {Cirac}}]{Kraus2010}%
  \BibitemOpen
  \bibfield  {author} {\bibinfo {author} {\bibfnamefont {C.~V.}\ \bibnamefont
  {Kraus}}, \bibinfo {author} {\bibfnamefont {N.}~\bibnamefont {Schuch}},
  \bibinfo {author} {\bibfnamefont {F.}~\bibnamefont {Verstraete}}, \ and\
  \bibinfo {author} {\bibfnamefont {J.~I.}\ \bibnamefont {Cirac}},\ }\href
  {\doibase 10.1103/PhysRevA.81.052338} {\bibfield  {journal} {\bibinfo
  {journal} {Phys. Rev. A}\ }\textbf {\bibinfo {volume} {81}},\ \bibinfo
  {pages} {052338} (\bibinfo {year} {2010})}\BibitemShut {NoStop}%
\bibitem [{\citenamefont {Wahl}\ \emph {et~al.}(2013)\citenamefont {Wahl},
  \citenamefont {Tu}, \citenamefont {Schuch},\ and\ \citenamefont
  {Cirac}}]{Wahl2013}%
  \BibitemOpen
  \bibfield  {author} {\bibinfo {author} {\bibfnamefont {T.~B.}\ \bibnamefont
  {Wahl}}, \bibinfo {author} {\bibfnamefont {H.-H.}\ \bibnamefont {Tu}},
  \bibinfo {author} {\bibfnamefont {N.}~\bibnamefont {Schuch}}, \ and\ \bibinfo
  {author} {\bibfnamefont {J.~I.}\ \bibnamefont {Cirac}},\ }\href {\doibase
  10.1103/PhysRevLett.111.236805} {\bibfield  {journal} {\bibinfo  {journal}
  {Phys. Rev. Lett.}\ }\textbf {\bibinfo {volume} {111}},\ \bibinfo {pages}
  {236805} (\bibinfo {year} {2013})}\BibitemShut {NoStop}%
\bibitem [{\citenamefont {Dubail}\ and\ \citenamefont
  {Read}(2015)}]{Dubail2015}%
  \BibitemOpen
  \bibfield  {author} {\bibinfo {author} {\bibfnamefont {J.}~\bibnamefont
  {Dubail}}\ and\ \bibinfo {author} {\bibfnamefont {N.}~\bibnamefont {Read}},\
  }\href {\doibase 10.1103/PhysRevB.92.205307} {\bibfield  {journal} {\bibinfo
  {journal} {Phys. Rev. B}\ }\textbf {\bibinfo {volume} {92}},\ \bibinfo
  {pages} {205307} (\bibinfo {year} {2015})}\BibitemShut {NoStop}%
\bibitem [{\citenamefont {Poilblanc}\ \emph {et~al.}(2014)\citenamefont
  {Poilblanc}, \citenamefont {Corboz}, \citenamefont {Schuch},\ and\
  \citenamefont {Cirac}}]{Poilblanc2014}%
  \BibitemOpen
  \bibfield  {author} {\bibinfo {author} {\bibfnamefont {D.}~\bibnamefont
  {Poilblanc}}, \bibinfo {author} {\bibfnamefont {P.}~\bibnamefont {Corboz}},
  \bibinfo {author} {\bibfnamefont {N.}~\bibnamefont {Schuch}}, \ and\ \bibinfo
  {author} {\bibfnamefont {J.~I.}\ \bibnamefont {Cirac}},\ }\href {\doibase
  10.1103/PhysRevB.89.241106} {\bibfield  {journal} {\bibinfo  {journal} {Phys.
  Rev. B}\ }\textbf {\bibinfo {volume} {89}},\ \bibinfo {pages} {241106}
  (\bibinfo {year} {2014})}\BibitemShut {NoStop}%
\bibitem [{\citenamefont {Wahl}\ \emph {et~al.}(2014)\citenamefont {Wahl},
  \citenamefont {Ha\ss{}ler}, \citenamefont {Tu}, \citenamefont {Cirac},\ and\
  \citenamefont {Schuch}}]{Wahl2014}%
  \BibitemOpen
  \bibfield  {author} {\bibinfo {author} {\bibfnamefont {T.~B.}\ \bibnamefont
  {Wahl}}, \bibinfo {author} {\bibfnamefont {S.~T.}\ \bibnamefont
  {Ha\ss{}ler}}, \bibinfo {author} {\bibfnamefont {H.-H.}\ \bibnamefont {Tu}},
  \bibinfo {author} {\bibfnamefont {J.~I.}\ \bibnamefont {Cirac}}, \ and\
  \bibinfo {author} {\bibfnamefont {N.}~\bibnamefont {Schuch}},\ }\href
  {\doibase 10.1103/PhysRevB.90.115133} {\bibfield  {journal} {\bibinfo
  {journal} {Phys. Rev. B}\ }\textbf {\bibinfo {volume} {90}},\ \bibinfo
  {pages} {115133} (\bibinfo {year} {2014})}\BibitemShut {NoStop}%
\bibitem [{\citenamefont {Yang}\ \emph {et~al.}(2015)\citenamefont {Yang},
  \citenamefont {Wahl}, \citenamefont {Tu}, \citenamefont {Schuch},\ and\
  \citenamefont {Cirac}}]{Yang2015}%
  \BibitemOpen
  \bibfield  {author} {\bibinfo {author} {\bibfnamefont {S.}~\bibnamefont
  {Yang}}, \bibinfo {author} {\bibfnamefont {T.~B.}\ \bibnamefont {Wahl}},
  \bibinfo {author} {\bibfnamefont {H.-H.}\ \bibnamefont {Tu}}, \bibinfo
  {author} {\bibfnamefont {N.}~\bibnamefont {Schuch}}, \ and\ \bibinfo {author}
  {\bibfnamefont {J.~I.}\ \bibnamefont {Cirac}},\ }\href {\doibase
  10.1103/PhysRevLett.114.106803} {\bibfield  {journal} {\bibinfo  {journal}
  {Phys. Rev. Lett.}\ }\textbf {\bibinfo {volume} {114}},\ \bibinfo {pages}
  {106803} (\bibinfo {year} {2015})}\BibitemShut {NoStop}%
\bibitem [{\citenamefont {Hackenbroich}\ \emph {et~al.}(2020)\citenamefont
  {Hackenbroich}, \citenamefont {Bernevig}, \citenamefont {Schuch},\ and\
  \citenamefont {Regnault}}]{Hackenbroich2020}%
  \BibitemOpen
  \bibfield  {author} {\bibinfo {author} {\bibfnamefont {A.}~\bibnamefont
  {Hackenbroich}}, \bibinfo {author} {\bibfnamefont {B.~A.}\ \bibnamefont
  {Bernevig}}, \bibinfo {author} {\bibfnamefont {N.}~\bibnamefont {Schuch}}, \
  and\ \bibinfo {author} {\bibfnamefont {N.}~\bibnamefont {Regnault}},\ }\href
  {\doibase 10.1103/PhysRevB.101.115134} {\bibfield  {journal} {\bibinfo
  {journal} {Phys. Rev. B}\ }\textbf {\bibinfo {volume} {101}},\ \bibinfo
  {pages} {115134} (\bibinfo {year} {2020})}\BibitemShut {NoStop}%
\bibitem [{\citenamefont {Mortier}\ \emph {et~al.}()\citenamefont {Mortier},
  \citenamefont {Schuch}, \citenamefont {Verstraete},\ and\ \citenamefont
  {Haegeman}}]{Mortier2020}%
  \BibitemOpen
  \bibfield  {author} {\bibinfo {author} {\bibfnamefont {Q.}~\bibnamefont
  {Mortier}}, \bibinfo {author} {\bibfnamefont {N.}~\bibnamefont {Schuch}},
  \bibinfo {author} {\bibfnamefont {F.}~\bibnamefont {Verstraete}}, \ and\
  \bibinfo {author} {\bibfnamefont {J.}~\bibnamefont {Haegeman}},\ }\href@noop
  {} {\ }\Eprint {http://arxiv.org/abs/arXiv:2008.11176} {arXiv:2008.11176}
  \BibitemShut {NoStop}%
\bibitem [{Note1()}]{Note1}%
  \BibitemOpen
  \bibinfo {note} {This can be straightforwardly generalized to the case of a
  larger unit cell, where different sites in the same unit cell have different
  $T_{\protect \mathbf {r}}$.}\BibitemShut {Stop}%
\bibitem [{\citenamefont {Peschel}(2003)}]{Peschel2003}%
  \BibitemOpen
  \bibfield  {author} {\bibinfo {author} {\bibfnamefont {I.}~\bibnamefont
  {Peschel}},\ }\href {https://doi.org/10.1088/0305-4470/36/14/101} {\bibfield
  {journal} {\bibinfo  {journal} {J. Phys. A}\ }\textbf {\bibinfo {volume}
  {36}},\ \bibinfo {pages} {L205} (\bibinfo {year} {2003})}\BibitemShut
  {NoStop}%
\bibitem [{\citenamefont {Bravyi}(2005)}]{Bravyi2005}%
  \BibitemOpen
  \bibfield  {author} {\bibinfo {author} {\bibfnamefont {S.}~\bibnamefont
  {Bravyi}},\ }\href {\doibase 10.26421/QIC5.3-3} {\bibfield  {journal}
  {\bibinfo  {journal} {Quantum Inf. and Comp.}\ }\textbf {\bibinfo {volume}
  {5}},\ \bibinfo {pages} {216} (\bibinfo {year} {2005})}\BibitemShut {NoStop}%
\bibitem [{\citenamefont {Wolf}(2006)}]{Wolf2006}%
  \BibitemOpen
  \bibfield  {author} {\bibinfo {author} {\bibfnamefont {M.~M.}\ \bibnamefont
  {Wolf}},\ }\href {\doibase 10.1103/PhysRevLett.96.010404} {\bibfield
  {journal} {\bibinfo  {journal} {Phys. Rev. Lett.}\ }\textbf {\bibinfo
  {volume} {96}},\ \bibinfo {pages} {010404} (\bibinfo {year}
  {2006})}\BibitemShut {NoStop}%
\bibitem [{\citenamefont {Gioev}\ and\ \citenamefont
  {Klich}(2006)}]{Gioev2006}%
  \BibitemOpen
  \bibfield  {author} {\bibinfo {author} {\bibfnamefont {D.}~\bibnamefont
  {Gioev}}\ and\ \bibinfo {author} {\bibfnamefont {I.}~\bibnamefont {Klich}},\
  }\href {\doibase 10.1103/PhysRevLett.96.100503} {\bibfield  {journal}
  {\bibinfo  {journal} {Phys. Rev. Lett.}\ }\textbf {\bibinfo {volume} {96}},\
  \bibinfo {pages} {100503} (\bibinfo {year} {2006})}\BibitemShut {NoStop}%
\bibitem [{\citenamefont {Edelman}\ \emph {et~al.}(1998)\citenamefont
  {Edelman}, \citenamefont {Arias},\ and\ \citenamefont {Smith}}]{Edelman1998}%
  \BibitemOpen
  \bibfield  {author} {\bibinfo {author} {\bibfnamefont {A.}~\bibnamefont
  {Edelman}}, \bibinfo {author} {\bibfnamefont {T.~A.}\ \bibnamefont {Arias}},
  \ and\ \bibinfo {author} {\bibfnamefont {S.~T.}\ \bibnamefont {Smith}},\
  }\href {\doibase 10.1137/s0895479895290954} {\bibfield  {journal} {\bibinfo
  {journal} {SIAM J. Matrix Anal. Appl.}\ }\textbf {\bibinfo {volume} {20}},\
  \bibinfo {pages} {303} (\bibinfo {year} {1998})}\BibitemShut {NoStop}%
\bibitem [{\citenamefont {Voorhis}\ and\ \citenamefont
  {Head-Gordon}(2002)}]{VOORHIS2002}%
  \BibitemOpen
  \bibfield  {author} {\bibinfo {author} {\bibfnamefont {T.~V.}\ \bibnamefont
  {Voorhis}}\ and\ \bibinfo {author} {\bibfnamefont {M.}~\bibnamefont
  {Head-Gordon}},\ }\href {\doibase 10.1080/00268970110103642} {\bibfield
  {journal} {\bibinfo  {journal} {Mol. Phys.}\ }\textbf {\bibinfo {volume}
  {100}},\ \bibinfo {pages} {1713} (\bibinfo {year} {2002})}\BibitemShut
  {NoStop}%
\bibitem [{\citenamefont {Abrudan}\ \emph {et~al.}(2009)\citenamefont
  {Abrudan}, \citenamefont {Eriksson},\ and\ \citenamefont
  {Koivunen}}]{Abrudan2009}%
  \BibitemOpen
  \bibfield  {author} {\bibinfo {author} {\bibfnamefont {T.}~\bibnamefont
  {Abrudan}}, \bibinfo {author} {\bibfnamefont {J.}~\bibnamefont {Eriksson}}, \
  and\ \bibinfo {author} {\bibfnamefont {V.}~\bibnamefont {Koivunen}},\ }\href
  {\doibase 10.1016/j.sigpro.2009.03.015} {\bibfield  {journal} {\bibinfo
  {journal} {Signal Process.}\ }\textbf {\bibinfo {volume} {89}},\ \bibinfo
  {pages} {1704} (\bibinfo {year} {2009})}\BibitemShut {NoStop}%
\bibitem [{\citenamefont {Zhu}(2016)}]{Zhu2016}%
  \BibitemOpen
  \bibfield  {author} {\bibinfo {author} {\bibfnamefont {X.}~\bibnamefont
  {Zhu}},\ }\href {\doibase 10.1007/s10589-016-9883-4} {\bibfield  {journal}
  {\bibinfo  {journal} {Comput. Optim. Appl.}\ }\textbf {\bibinfo {volume}
  {67}},\ \bibinfo {pages} {73} (\bibinfo {year} {2016})}\BibitemShut {NoStop}%
\bibitem [{\citenamefont {Hauru}\ \emph {et~al.}(2021)\citenamefont {Hauru},
  \citenamefont {Damme},\ and\ \citenamefont {Haegeman}}]{Hauru2021}%
  \BibitemOpen
  \bibfield  {author} {\bibinfo {author} {\bibfnamefont {M.}~\bibnamefont
  {Hauru}}, \bibinfo {author} {\bibfnamefont {M.~V.}\ \bibnamefont {Damme}}, \
  and\ \bibinfo {author} {\bibfnamefont {J.}~\bibnamefont {Haegeman}},\ }\href
  {\doibase 10.21468/SciPostPhys.10.2.040} {\bibfield  {journal} {\bibinfo
  {journal} {SciPost Phys.}\ }\textbf {\bibinfo {volume} {10}},\ \bibinfo
  {pages} {40} (\bibinfo {year} {2021})}\BibitemShut {NoStop}%
\bibitem [{\citenamefont {Nocedal}\ and\ \citenamefont
  {Wright}(2006)}]{Nocedal2006}%
  \BibitemOpen
  \bibfield  {author} {\bibinfo {author} {\bibfnamefont {J.}~\bibnamefont
  {Nocedal}}\ and\ \bibinfo {author} {\bibfnamefont {S.~J.}\ \bibnamefont
  {Wright}},\ }\href {\doibase 10.1007/978-0-387-40065-5} {\emph {\bibinfo
  {title} {Numerical Optimization}}}\ (\bibinfo  {publisher} {Springer New
  York},\ \bibinfo {year} {2006})\BibitemShut {NoStop}%
\bibitem [{\citenamefont {Cs{\'{a}}sz{\'{a}}r}\ and\ \citenamefont
  {Pulay}(1984)}]{Csaszar1984}%
  \BibitemOpen
  \bibfield  {author} {\bibinfo {author} {\bibfnamefont {P.}~\bibnamefont
  {Cs{\'{a}}sz{\'{a}}r}}\ and\ \bibinfo {author} {\bibfnamefont
  {P.}~\bibnamefont {Pulay}},\ }\href {\doibase 10.1016/s0022-2860(84)87198-7}
  {\bibfield  {journal} {\bibinfo  {journal} {J. Mol. Struct.}\ }\textbf
  {\bibinfo {volume} {114}},\ \bibinfo {pages} {31} (\bibinfo {year}
  {1984})}\BibitemShut {NoStop}%
\bibitem [{\citenamefont {Corboz}\ \emph {et~al.}(2010)\citenamefont {Corboz},
  \citenamefont {Or\'us}, \citenamefont {Bauer},\ and\ \citenamefont
  {Vidal}}]{Corboz2010}%
  \BibitemOpen
  \bibfield  {author} {\bibinfo {author} {\bibfnamefont {P.}~\bibnamefont
  {Corboz}}, \bibinfo {author} {\bibfnamefont {R.}~\bibnamefont {Or\'us}},
  \bibinfo {author} {\bibfnamefont {B.}~\bibnamefont {Bauer}}, \ and\ \bibinfo
  {author} {\bibfnamefont {G.}~\bibnamefont {Vidal}},\ }\href {\doibase
  10.1103/PhysRevB.81.165104} {\bibfield  {journal} {\bibinfo  {journal} {Phys.
  Rev. B}\ }\textbf {\bibinfo {volume} {81}},\ \bibinfo {pages} {165104}
  (\bibinfo {year} {2010})}\BibitemShut {NoStop}%
\bibitem [{\citenamefont {Weichselbaum}(2012)}]{Weichselbaum2012}%
  \BibitemOpen
  \bibfield  {author} {\bibinfo {author} {\bibfnamefont {A.}~\bibnamefont
  {Weichselbaum}},\ }\href {\doibase https://doi.org/10.1016/j.aop.2012.07.009}
  {\bibfield  {journal} {\bibinfo  {journal} {Ann. Phys.}\ }\textbf {\bibinfo
  {volume} {327}},\ \bibinfo {pages} {2972} (\bibinfo {year}
  {2012})}\BibitemShut {NoStop}%
\bibitem [{\citenamefont {Weichselbaum}(2020)}]{Weichselbaum2020}%
  \BibitemOpen
  \bibfield  {author} {\bibinfo {author} {\bibfnamefont {A.}~\bibnamefont
  {Weichselbaum}},\ }\href {\doibase 10.1103/PhysRevResearch.2.023385}
  {\bibfield  {journal} {\bibinfo  {journal} {Phys. Rev. Research}\ }\textbf
  {\bibinfo {volume} {2}},\ \bibinfo {pages} {023385} (\bibinfo {year}
  {2020})}\BibitemShut {NoStop}%
\bibitem [{\citenamefont {Affleck}\ and\ \citenamefont
  {Marston}(1988)}]{Affleck1988}%
  \BibitemOpen
  \bibfield  {author} {\bibinfo {author} {\bibfnamefont {I.}~\bibnamefont
  {Affleck}}\ and\ \bibinfo {author} {\bibfnamefont {J.~B.}\ \bibnamefont
  {Marston}},\ }\href {\doibase 10.1103/PhysRevB.37.3774} {\bibfield  {journal}
  {\bibinfo  {journal} {Phys. Rev. B}\ }\textbf {\bibinfo {volume} {37}},\
  \bibinfo {pages} {3774} (\bibinfo {year} {1988})}\BibitemShut {NoStop}%
\bibitem [{\citenamefont {Hastings}(2000)}]{Hastings2000}%
  \BibitemOpen
  \bibfield  {author} {\bibinfo {author} {\bibfnamefont {M.~B.}\ \bibnamefont
  {Hastings}},\ }\href {\doibase 10.1103/PhysRevB.63.014413} {\bibfield
  {journal} {\bibinfo  {journal} {Phys. Rev. B}\ }\textbf {\bibinfo {volume}
  {63}},\ \bibinfo {pages} {014413} (\bibinfo {year} {2000})}\BibitemShut
  {NoStop}%
\bibitem [{\citenamefont {Ran}\ \emph {et~al.}(2007)\citenamefont {Ran},
  \citenamefont {Hermele}, \citenamefont {Lee},\ and\ \citenamefont
  {Wen}}]{Ran2007}%
  \BibitemOpen
  \bibfield  {author} {\bibinfo {author} {\bibfnamefont {Y.}~\bibnamefont
  {Ran}}, \bibinfo {author} {\bibfnamefont {M.}~\bibnamefont {Hermele}},
  \bibinfo {author} {\bibfnamefont {P.~A.}\ \bibnamefont {Lee}}, \ and\
  \bibinfo {author} {\bibfnamefont {X.-G.}\ \bibnamefont {Wen}},\ }\href
  {\doibase 10.1103/PhysRevLett.98.117205} {\bibfield  {journal} {\bibinfo
  {journal} {Phys. Rev. Lett.}\ }\textbf {\bibinfo {volume} {98}},\ \bibinfo
  {pages} {117205} (\bibinfo {year} {2007})}\BibitemShut {NoStop}%
\bibitem [{\citenamefont {Hermele}\ \emph {et~al.}(2008)\citenamefont
  {Hermele}, \citenamefont {Ran}, \citenamefont {Lee},\ and\ \citenamefont
  {Wen}}]{Hermele2008}%
  \BibitemOpen
  \bibfield  {author} {\bibinfo {author} {\bibfnamefont {M.}~\bibnamefont
  {Hermele}}, \bibinfo {author} {\bibfnamefont {Y.}~\bibnamefont {Ran}},
  \bibinfo {author} {\bibfnamefont {P.~A.}\ \bibnamefont {Lee}}, \ and\
  \bibinfo {author} {\bibfnamefont {X.-G.}\ \bibnamefont {Wen}},\ }\href
  {\doibase 10.1103/PhysRevB.77.224413} {\bibfield  {journal} {\bibinfo
  {journal} {Phys. Rev. B}\ }\textbf {\bibinfo {volume} {77}},\ \bibinfo
  {pages} {224413} (\bibinfo {year} {2008})}\BibitemShut {NoStop}%
\bibitem [{\citenamefont {Hermele}\ \emph {et~al.}(2004)\citenamefont
  {Hermele}, \citenamefont {Senthil}, \citenamefont {Fisher}, \citenamefont
  {Lee}, \citenamefont {Nagaosa},\ and\ \citenamefont {Wen}}]{Hermele2004}%
  \BibitemOpen
  \bibfield  {author} {\bibinfo {author} {\bibfnamefont {M.}~\bibnamefont
  {Hermele}}, \bibinfo {author} {\bibfnamefont {T.}~\bibnamefont {Senthil}},
  \bibinfo {author} {\bibfnamefont {M.~P.~A.}\ \bibnamefont {Fisher}}, \bibinfo
  {author} {\bibfnamefont {P.~A.}\ \bibnamefont {Lee}}, \bibinfo {author}
  {\bibfnamefont {N.}~\bibnamefont {Nagaosa}}, \ and\ \bibinfo {author}
  {\bibfnamefont {X.-G.}\ \bibnamefont {Wen}},\ }\href {\doibase
  10.1103/PhysRevB.70.214437} {\bibfield  {journal} {\bibinfo  {journal} {Phys.
  Rev. B}\ }\textbf {\bibinfo {volume} {70}},\ \bibinfo {pages} {214437}
  (\bibinfo {year} {2004})}\BibitemShut {NoStop}%
\bibitem [{\citenamefont {Ivanov}(1999)}]{Ivanov1999}%
  \BibitemOpen
  \bibfield  {author} {\bibinfo {author} {\bibfnamefont {D.~A.}\ \bibnamefont
  {Ivanov}},\ }\emph {\bibinfo {title} {On the SU$(2)$ theory of the $t$-$J$
  model}},\ \href {https://dspace.mit.edu/handle/1721.1/84774} {Ph.D. thesis},\
  \bibinfo  {school} {MIT} (\bibinfo {year} {1999})\BibitemShut {NoStop}%
\bibitem [{\citenamefont {Ferrari}\ \emph {et~al.}(2021)\citenamefont
  {Ferrari}, \citenamefont {Parola},\ and\ \citenamefont
  {Becca}}]{Ferrari2021}%
  \BibitemOpen
  \bibfield  {author} {\bibinfo {author} {\bibfnamefont {F.}~\bibnamefont
  {Ferrari}}, \bibinfo {author} {\bibfnamefont {A.}~\bibnamefont {Parola}}, \
  and\ \bibinfo {author} {\bibfnamefont {F.}~\bibnamefont {Becca}},\ }\href
  {\doibase 10.1103/PhysRevB.103.195140} {\bibfield  {journal} {\bibinfo
  {journal} {Phys. Rev. B}\ }\textbf {\bibinfo {volume} {103}},\ \bibinfo
  {pages} {195140} (\bibinfo {year} {2021})}\BibitemShut {NoStop}%
\bibitem [{\citenamefont {Rantner}\ and\ \citenamefont
  {Wen}(2002)}]{Rantner2002}%
  \BibitemOpen
  \bibfield  {author} {\bibinfo {author} {\bibfnamefont {W.}~\bibnamefont
  {Rantner}}\ and\ \bibinfo {author} {\bibfnamefont {X.-G.}\ \bibnamefont
  {Wen}},\ }\href {\doibase 10.1103/PhysRevB.66.144501} {\bibfield  {journal}
  {\bibinfo  {journal} {Phys. Rev. B}\ }\textbf {\bibinfo {volume} {66}},\
  \bibinfo {pages} {144501} (\bibinfo {year} {2002})}\BibitemShut {NoStop}%
\bibitem [{\citenamefont {Yang}\ \emph {et~al.}()\citenamefont {Yang},
  \citenamefont {Zhang}, \citenamefont {Liao}, \citenamefont {Tu},\ and\
  \citenamefont {Wang}}]{Yang2022}%
  \BibitemOpen
  \bibfield  {author} {\bibinfo {author} {\bibfnamefont {Q.}~\bibnamefont
  {Yang}}, \bibinfo {author} {\bibfnamefont {X.-Y.}\ \bibnamefont {Zhang}},
  \bibinfo {author} {\bibfnamefont {H.-J.}\ \bibnamefont {Liao}}, \bibinfo
  {author} {\bibfnamefont {H.-H.}\ \bibnamefont {Tu}}, \ and\ \bibinfo {author}
  {\bibfnamefont {L.}~\bibnamefont {Wang}},\ }\href@noop {} {\ }\Eprint
  {http://arxiv.org/abs/arXiv:2208.xxxxx} {arXiv:2208.xxxxx} \BibitemShut
  {NoStop}%
\end{thebibliography}%


%
\bibliographystyle{apsrev4-1}

\end{document}